\documentclass[letterpaper,reprint,superscriptaddress,twocolumn,aps,prb,showpacs,preprintnumbers]{revtex4}
\usepackage{graphicx}  
\usepackage{bm}        
\usepackage{amssymb}   
\usepackage{amsmath}   

\usepackage[colorlinks,hyperindex]{hyperref}
 \hypersetup
 {
     urlcolor=blue,%
 }

\usepackage{color}

\newcommand{\up}{\uparrow}
\newcommand{\dw}{\downarrow}

\begin{document}

\title{Mass-Imbalanced Ionic Hubbard Chain}
\author{Michael Sekania}
\affiliation{Theoretical Physics III, Center for Electronic Correlations and Magnetism, Institute of Physics, \\
             University of Augsburg, D-86135 Augsburg, Germany
            }
\affiliation{Javakhishvili Tbilisi State University,
             Andronikashvili Institute of Physics,
             Tamarashvili str.~6, 0177 Tbilisi, Georgia
            }

\author{Dionys Baeriswyl}
\affiliation{Department of Physics,
             University of Fribourg,
             Chemin du Mus\'ee 3, CH-1700 Fribourg, Switzerland
            }
\affiliation{International Institute of Physics,
             59078-400 Natal-RN, Brazil
            }

\author{Luka Jibuti}
\affiliation{Faculty of Natural Sciences and Engineering,
             Ilia State University,
             Cholokashvili Avenue 3-5, 0162 Tbilisi, Georgia
            }

\author{George I. Japaridze}
\affiliation{Javakhishvili Tbilisi State University,
             Andronikashvili Institute of Physics,
             Tamarashvili str.~6, 0177 Tbilisi, Georgia
            }
\affiliation{Faculty of Natural Sciences and Engineering,
             Ilia State University,
             Cholokashvili Avenue 3-5, 0162 Tbilisi, Georgia
            }

\begin{abstract}
A repulsive Hubbard model with both spin-asymmetric hopping (${t_\uparrow\neq t_\downarrow}$) and a staggered potential (of strength $\Delta$) is studied in one dimension.
The model is a compound of the mass-imbalanced  (${t_\uparrow\neq t_\downarrow}$, ${\Delta=0}$) and ionic  (${t_\uparrow = t_\downarrow}$, ${\Delta>0}$) Hubbard models, and may be realized by cold atoms in engineered optical lattices.
We use mostly mean-field theory to determine the phases and phase transitions in the ground state for a half-filled band (one particle per site).
We find that a period-two modulation of the particle (or charge) density  and an alternating spin density coexist for arbitrary Hubbard interaction strength, ${U\geqslant 0}$.
The amplitude of the charge modulation is largest at ${U=0}$, decreases with increasing $U$ and tends to zero for ${U\rightarrow\infty}$.
The amplitude for spin alternation increases with $U$ and tends to saturation for ${U\rightarrow\infty}$.
Charge order dominates below a value $U_c$, whereas magnetic order dominates above.
The mean-field Hamiltonian has two gap parameters, $\Delta_\uparrow$ and $\Delta_\downarrow$, which have to be determined self-consistently.
For ${U<U_c}$ both parameters are positive, for ${U>U_c}$ they have different signs, and for ${U=U_c}$ one gap parameter jumps from a positive to a negative value.
The weakly first-order phase transition at $U_c$ can be interpreted in terms of an avoided criticality (or metallicity).
The system is reluctant  to restore a symmetry that has been broken explicitly.
\end{abstract}


\maketitle
\section{Introduction} \label{sec:introduction}
During the last decade, the study of ultracold atoms in optical lattices has spawned new insight into the complex behavior of quantum many-body systems 
[\onlinecite{Lewenstein_07,UCA-Review-MP1,UCA-Review-MP2}]. Atomic gases stored in 
artificially engineered optical lattices constitute structures beyond those
currently achievable in actual materials and, thanks to the easy manipulation of parameters, serve as a playground for the simulation of condensed-matter systems  with
unconventional properties  [\onlinecite{Wilczek04}].  While in the solid-state context the use of simplified models cannot always be justified, because they may neglect relevant degrees of freedom, the
clean and precisely controlled environment of ultracold atoms in optical
lattices allows a direct mapping of a physical reality onto such models. Moreover,
the possibility of manipulating the interaction strength using the Feshbach
resonance [\onlinecite{UCA_Feshbach_Rev}] enables the observation of many-body phenomena from weak to strong coupling. Both
bosonic [\onlinecite{Jaksch_98,Greiner_02}] and fermionic
[\onlinecite{Hofstetter_02,Esslinger_08,Esslinger_10}] Hubbard models have been
realized and investigated experimentally.

Optical lattices can be generated in various geometries, including bipartite lattices with different potential minima on the two sublattices [\onlinecite{Hemmerich_91}].  For bosonic atoms a checkerboard potential has been used to study metastable Bose-Einstein condensates with unconventional order parameters
[\onlinecite{Olschlager_11, Olschlager_13, Diliberto_14}]. The relaxation of a bosonic gas, initially prepared in a state with alternating site occupancies, has been investigated for a one-dimensional optical lattice, by switching off and on a period-two potential [\onlinecite{Trotzky_12}]. Fermionic atomic gases in an optical honeycomb lattice -- artificial graphene -- have been tuned to the state of a Mott insulator for large enough on-site coupling $U$ [\onlinecite{Tarruel_12}]. With the addition of a disparity of well depths on the $A$ and $B$ sites of the honeycomb lattice a competition between the Mott insulator with homogeneous particle density and a band insulator with a modulated density appears [\onlinecite{Esslinger_13, Messer_15}]. In the limit of deep wells, such a system can be described by the ionic Hubbard model, where the single-particle levels on neighboring sites differ by an energy $\Delta$. In condensed-matter physics this model has appeared a long time ago in the context of the neutral-to-ionic transition [\onlinecite{Nagaosa_86}], observed in organic mixed-stack compounds [\onlinecite{Torrance_81, Hubbard_81}]. Later similar models have been used for describing metal-halogen chains [\onlinecite{Gammel_90}] and transition-metal oxides [\onlinecite{Egami_93,Egami_94}]. In these cases the $A$ and $B$ sites are associated with $d$- and $p$-orbitals, respectively. 

The conventional Hubbard Hamiltonian is invariant with respect to translations by arbitrary lattice vectors and also has spin SU(2) symmetry. In the ionic Hubbard model the translational symmetry is reduced -- the unit cell contains two sites -- but the Hamiltonian preserves the SU(2) invariance. SU(2) symmetry can be broken in more than one way. One option for electronic systems is the coupling to a magnetic field, which induces an imbalance between up and down spins. In systems of cold atoms, where the spin degree of freedom may represent two hyperfine levels, an uneven mixture of the two components leads to a similar population imbalance [\onlinecite{Zwierlein_06}]. Another option for cold atoms is to use two atom species with different masses 
[\onlinecite{Wille-et-al-08,Taglieber-et-al-08}]. This mass imbalance leads to different hopping parameters $t_\uparrow$ and $t_\downarrow$ in the Hubbard model. State-dependent tunneling has also been realized for gases with one atomic species in two hyperfine levels [\onlinecite{Mandel_04,Messer_15a}].  
In condensed-matter physics mixtures of fermions with different effective masses are realized in rare-earth-metal compounds where a localized $f$ level crosses a wide conduction band [\onlinecite{Fazekas_99, Riseborough_16}]. However, in the models advocated for describing these materials the two bands are usually strongly hybridized, and at least the light electrons have an additional spin degree of freedom. 

In this paper we study the mass-imbalanced ionic Hubbard model, where translational and spin SU(2) symmetries are both explicitly broken. The Hamiltonian and the order parameters are presented in Sec.~\ref{sec:hamiltonian}. In Sec.~\ref{sec:limits} several limits are discussed, including the non-interacting case (${U=0}$), the small- and large-$U$ limits, the conventional ionic Hubbard model 
(${t_\uparrow=t_\downarrow}$, ${\Delta > 0}$) and the mass-imbalanced Hubbard model (${t_\uparrow \neq t_\downarrow}$, ${\Delta = 0}$).
In Secs.~\ref{sec:mft} and \ref{sec:pt} we concentrate on mean-field predictions for the ground state. The method is explained in Sec.~\ref{sec:mft} and used in Sec.~\ref{sec:pt} for describing the phase transition from (dominant) charge to spin order. It is found that the phase diagram for the mass-imbalanced case differs qualitatively from that of the conventional ionic Hubbard model. A brief summary is presented in Sec.~\ref{sec:sum}, which also  lists some questions to be addressed in the future.

\section{Hamiltonian and order parameters}\label{sec:hamiltonian}
The mass-imbalanced ionic Hubbard chain is defined by the Hamiltonian 
\begin{align}
H =&-\sum_{i\sigma}t_\sigma (c_{i\sigma}^\dag c_{i+1\sigma}^{\phantom{}} +c_{i+1\sigma}^\dag c_{i\sigma}^{\phantom{}}) \nonumber \\
   &+\frac{\Delta}{2}\sum_{i\sigma}(-1)^in_{i\sigma} +U\sum_in_{i\uparrow}n_{i\downarrow},
\label{eq:ham1}
\end{align}
where $c_{i\sigma}^\dag$ and $c_{i\sigma}^{\phantom{}}$ create and annihilate, respectively, fermionic particles at sites $i$ ($i=1,\dots,L$) with spin projections ${\sigma=\uparrow,\downarrow}$,
and ${n_{i\sigma}=c_{i\sigma}^\dag c_{i\sigma}^{\phantom{}}}$.
The Hamiltonian commutes with the number of particles with spin $\sigma$, ${N_\sigma=\sum_in_{i\sigma}}$, and hence with the total particle number, ${N=N_\uparrow+N_\downarrow}$. We restrict ourselves to the case of half filling, where $N$ is equal to the number of sites $L$. Moreover, we assume that the total magnetization vanishes, i.e., ${N_\uparrow=N_\downarrow}$. The parameters of the Hamiltonian will be chosen in the range 
${t_\uparrow\geqslant t_\downarrow}$, ${\Delta\geqslant 0}$ and ${U\geqslant 0}$. 

A finite ionic term (${\Delta>0}$) breaks translational invariance and induces a density imbalance between neighboring sites, whereas the Hubbard term (${U>0}$) suppresses density inhomogeneities and favors antiferromagnetic ordering. For different values of the hopping amplitudes (${t_\uparrow\neq t_\downarrow}$) the spin SU(2) symmetry is also explicitly broken. In this case charge and spin modulations are expected to be non-zero for any finite values of $U$ and 
$\Delta$. 
To characterize the two broken symmetries we introduce the order parameters 
$\delta\rho_c$ and $\delta\rho_s$, defined by the relations
\begin{align}
\begin{split}
\delta\rho_c&=-\frac{1}{L}\sum_{i\sigma}(-1)^i n_{i\sigma},\\
\delta\rho_s&=\,\frac{1}{L}\sum_{i\sigma}(-1)^i\sigma n_{i\sigma}.
\end{split}
\label{eq:op1}
\end{align}
For ${\vert\delta\rho_c\vert>\vert\delta\rho_s\vert}$ charge ordering dominates (ionic phase), while for 
${\vert\delta\rho_s\vert>\vert\delta\rho_c\vert}$ spin ordering prevails (antiferromagnetic phase).

The canonical transformation
\begin{align}
c_{i\uparrow}^{\phantom{}}\rightarrow c_{i\uparrow}^{\phantom{}}\, ,\qquad c_{i\downarrow}^{\phantom{}}\rightarrow (-1)^{i}c_{i\downarrow}^\dag
\label{eq:mapping}
\end{align}
leaves the hopping term of the Hamiltonian \eqref{eq:ham1} invariant, but exchanges the order parameters, 
${\delta\rho_c\leftrightarrow -\delta\rho_s}$. The ionic potential is replaced by a Zeeman coupling to an alternating magnetic field and the interaction term changes sign. Therefore this transformation maps the repulsive ionic Hubbard model onto the attractive Hubbard model in an alternating magnetic field. The phase diagram for positive $U$ is then readily  converted into the phase diagram for negative $U$ by interchanging ionic and antiferromagnetic phases and by associating the parameter $\Delta$ with the amplitude of a staggered magnetic field. It has been argued that the mapping \eqref{eq:mapping} can be very useful for understanding the repulsive Hubbard model by using cold atoms in optical lattices as quantum simulators in the attractive regime [\onlinecite{Ho_09}].

The in-plane alternating spin 
$(-1)^ic_{i\uparrow}^\dag c_{i\downarrow}^{\phantom{}}$ is transformed by Eq.~\eqref{eq:mapping} to a pair operator
$c_{i\uparrow}^\dag c_{i\downarrow}^\dag$,
which implies an intimate relationship between in-plane antiferromagnetic ordering for the repulsive ionic Hubbard model and superconductivity for the attractive Hubbard model in an alternating magnetic field. This field is detrimental for superconductivity and thus favors an alternating charge density for the attractive Hubbard model. Correspondingly, we expect in-plane antiferromagnetism to be suppressed by the ionic potential, and this is indeed found both within mean-field theory and in the large $U$ limit. 

It is convenient to introduce unit cells with two sites and operators
\begin{align}
a_{m\sigma}=c_{2m-1\sigma},\quad b_{m\sigma}=c_{2m\sigma}, \quad m=1,\dots,\frac{L}{2}\,.
\end{align}
The Hamiltonian \eqref{eq:ham1} then reads
\begin{align}
H=&-\sum_{m\sigma}t_\sigma (a_{m\sigma}^\dag b_{m\sigma}^{\phantom{}}+b_{m\sigma}^\dag a_{m+1\sigma}^{\phantom{}}+\mbox{H.c.}) \nonumber \\
  &-\frac{\Delta}{2}\sum_{m\sigma}(n_{m\sigma}^A-n_{m\sigma}^B) \nonumber \\
  &+U\sum_m\left(n_{m\uparrow}^An_{m\downarrow}^A+n_{m\uparrow}^Bn_{m\downarrow}^B\right),
\label{eq:ham2}
\end{align}
where $n_{m\sigma}^A=a_{m\sigma}^\dag a_{m\sigma}^{\phantom{}},\, n_{m\sigma}^B=b_{m\sigma}^\dag b_{m\sigma}^{\phantom{}}$. 

\section{Limiting cases}\label{sec:limits}
Before embarking on a discussion of the mean-field ground state of the Hamiltonian \eqref{eq:ham2}, we explore certain limiting cases, namely $U=0$, small and large $U$, ${t_\uparrow=t_\downarrow}$, and ${\Delta=0}$.
\subsection{Non-interacting particles: $U=0$}\label{sec:nonint}
To diagonalize the Hamiltonian \eqref{eq:ham2} for ${U=0}$, we first represent the Wannier operators
$a_{m\sigma}, b_{m\sigma}$ by Bloch operators $a_{k\sigma}, b_{k\sigma}$,
\begin{align} 
\begin{split}
a_{m\sigma}^{\phantom{}}&=\sqrt{\frac{2}{L}}\sum_ke^{ikm}a_{k\sigma}^{\phantom{}}, \\
b_{m\sigma}^{\phantom{}}&=\sqrt{\frac{2}{L}}\sum_ke^{ik(m+\frac{1}{2})}b_{k\sigma}^{\phantom{}},
\end{split}
\end{align} 
where $k=\frac{4\pi}{N}\nu$, ${-\frac{L}{4}<\nu\leqslant\frac{L}{4}}$. The Hamiltonian then reads
\begin{align}
H_0=\sum_{k\sigma}\left(a_{k\sigma}^\dag,\, b_{k\sigma}^\dag\right)
\left(\begin{array}{cc}-\Delta/2&\varepsilon_{k\sigma}\\\varepsilon_{k\sigma}&\Delta/2\end{array}\right)
\left(\begin{array}{c}a_{k\sigma}^{\phantom{}}\\b_{k\sigma}^{\phantom{}}\end{array}\right),
\end{align}
where 
\begin{align}
\varepsilon_{k\sigma}^{\phantom{}}=-2t_\sigma \cos\frac{k}{2}\,.
\end{align}
The Bogoliubov transformation
\begin{align}
\begin{split}
a_{k\sigma}^{\phantom{}}&=\phantom{-}\cos \varphi_{k\sigma}^{\phantom{}}\alpha_{k\sigma}^{\phantom{}}+\sin \varphi_{k\sigma}^{\phantom{}}\beta_{k\sigma}^{\phantom{}},\\
b_{k\sigma}^{\phantom{}}&=-\sin \varphi_{k\sigma}^{\phantom{}}\alpha_{k\sigma}^{\phantom{}}+\cos \varphi_{k\sigma}^{\phantom{}}\beta_{k\sigma}^{\phantom{}},
\end{split}
\label{eq:bog}
\end{align}
diagonalizes $H_0$ if the angles $\varphi_{k\sigma}$ are chosen as
\begin{align}
\tan 2\varphi_{k\sigma}=\frac{2\varepsilon_{k\sigma}}{\Delta},\qquad
\cos 2\varphi_{k\sigma}=\frac{\Delta}{2E_{k\sigma}},
\label{eq:angles}
\end{align}
where
\begin{align}
E_{k\sigma}=\sqrt{\varepsilon_{k\sigma}^2+(\Delta/2)^2}\,.
\label{eq:spec}
\end{align}
The transformed Hamiltonian
\begin{align}
H_0=\sum_{k\sigma}E_{k\sigma}\left(-\alpha_{k\sigma}^\dag\alpha_{k\sigma}^{\phantom{}}
+\beta_{k\sigma}^\dag\beta_{k\sigma}^{\phantom{}}\right)
\label{eq:trham}
\end{align}
has conduction and valence bands 
separated from each other by an energy gap $\Delta$. For an average number of one particle per site, the case considered here, the occupation numbers in the ground state are 
${\langle\alpha_{k\sigma}^\dag\alpha_{k\sigma}^{\phantom{}}\rangle=1}$ and 
${\langle\beta_{k\sigma}^\dag\beta_{k\sigma}^{\phantom{}}\rangle=0}$, 
as in a conventional semiconductor.

For this ground state the order parameters \eqref{eq:op1} are easily evaluated in the thermodynamic limit, 
${L\rightarrow\infty}$, where $\frac{2}{L}\sum_k$ is replaced by $\frac{1}{2\pi}\int_{-\pi}^\pi dk$. We find
\begin{align}
\begin{split}
\delta\rho_c&=\,\frac{1}{L}\sum_{k\sigma}\cos 2\varphi_{k\sigma}
\rightarrow\,\frac{\Delta}{4\pi}\sum_\sigma\frac{\kappa_\sigma K(\kappa_\sigma)}{t_\sigma}, \\
\delta\rho_s&=-\frac{1}{L}\sum_{k\sigma}\sigma\cos 2\varphi_{k\sigma}
\rightarrow-\frac{\Delta}{4\pi}\sum_\sigma\sigma\frac{\kappa_\sigma K(\kappa_\sigma)}{t_\sigma},
\end{split}
\label{eq:ops}
\end{align}
where $K(\kappa_\sigma)$ is the complete elliptic integral of the first kind with modulus
\begin{align}
\kappa_\sigma=\left[1+\left(\frac{\Delta}{4t_\sigma}\right)^2\right]^{-\frac{1}{2}}.
\end{align}

\begin{figure}
  \centering
   \includegraphics[width=\columnwidth]{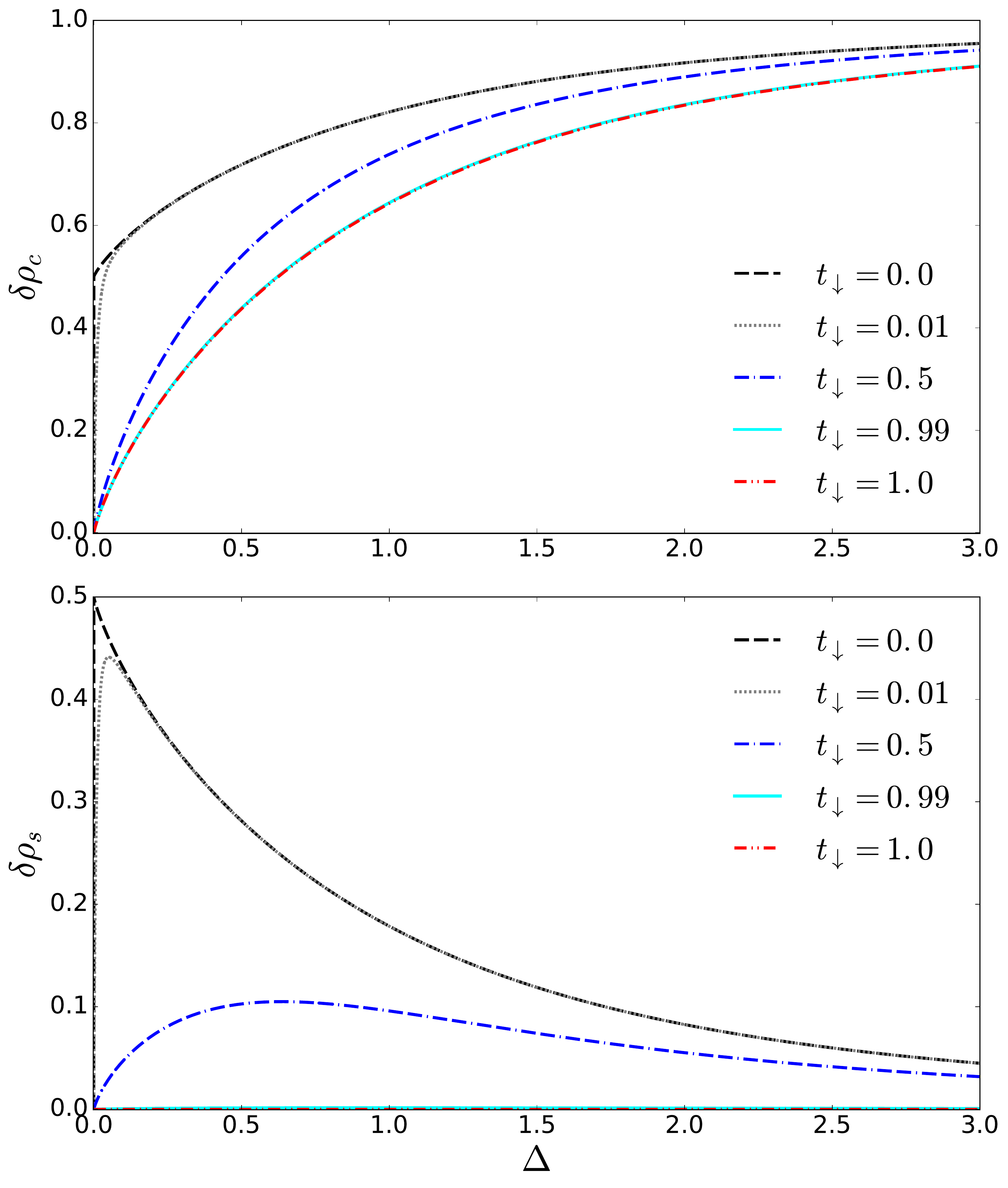}
   \caption{Charge modulation $\delta\rho$ and spin alternation $\rho_s$ as  functions of the ionicity parameter $\Delta$, for ${t_{\up}=1}$  and various values of the parameter $t_{\dw}$. The special case ${t_{\dw}=0}$ corresponds to the Falicov-Kimball limit.}
  \label{fig:1_FKM}
\end{figure}

Figure~\ref{fig:1_FKM} shows the amplitudes of both charge and spin modulations as functions of $\Delta$ for various ratios of hopping parameters. The limiting cases ${\Delta\rightarrow 0}$ and ${\Delta\rightarrow\infty}$ are readily obtained from Eq.~\eqref{eq:ops} using the asymptotic behavior of the elliptic integral,
\begin{align}
K(\kappa)\sim
  \left\{
    \begin{array}{ll}
      \frac{\pi}{2}\left[1+\frac{1}{4}\kappa^2+\frac{9}{64}\kappa^4+{\cal O}(\kappa^6)\right], & \kappa\rightarrow 0\,, \\ \\
      \frac{1}{2}\log\frac{16}{1-\kappa^2}, &\kappa\rightarrow 1\,.
    \end{array}
  \right.
\end{align}
If both $t_\uparrow$ and $t_\downarrow$ are finite $\delta\rho_c$ vanishes for ${\Delta\rightarrow 0}$ and tends to 1 for ${\Delta\rightarrow\infty}$, while $\delta\rho_s$ vanishes in both limits. The result for ${\Delta\rightarrow 0}$ is different if one of the hopping amplitudes, say $t_\downarrow$, vanishes (as in the Falicov-Kimball model). In this case both 
$\delta\rho_c$ and $\delta\rho_s$ tend to $\frac{1}{2}$ for ${\Delta\rightarrow 0}$. However, this is a quite singular limit. In fact, for ${\Delta=0}$ (and ${U=0}$) the total energy is completely independent of the distribution of the infinitely heavy particles. This degeneracy is removed by a finite value of $U$.

\subsection{Small and large $U$ limits}
For ${U\ll\Delta}$ we can use perturbation theory for calculating interaction effects. This is justified, because for ${U=0}$ there is a gap $\Delta$ in the excitation spectrum. To leading order in $U$ we find (for ${t_\uparrow\geqslant t_\downarrow}$)
\begin{align}
\delta\rho_c(U)&=\delta\rho_c(0) -\frac{\kappa_\uparrow\kappa_\downarrow U\Delta}{8\pi^2t_\uparrow t_\downarrow}\times \nonumber \\
               &\times\left[2K(\kappa_\uparrow)K(\kappa_\downarrow)
                -K(\kappa_\uparrow)E(\kappa_\downarrow) -E(\kappa_\uparrow)K(\kappa_\downarrow)\right] ,\nonumber\\
\delta\rho_s(U)&=\delta\rho_s(0) +\frac{\kappa_\uparrow\kappa_\downarrow U\Delta}{8\pi^2t_\uparrow t_\downarrow}\times \nonumber \\
               &\times\left[K(\kappa_\uparrow)E(\kappa_\downarrow)-E(\kappa_\uparrow)K(\kappa_\downarrow)\right] ,
\label{eq:pert}
\end{align}
where $E(\kappa_\sigma)$ is the complete elliptic integral of the second kind. The density modulation $\delta\rho_c$ decreases linearly with $U$, since ${E(x)<K(x)}$ for ${x>0}$. For our choice of hopping parameters 
we have ${\kappa_\uparrow\geqslant \kappa_\downarrow}$. The function $K(x)$ is monotonically increasing, while $E(x)$ decreases with $x$. Therefore ${\delta\rho_s(U)-\delta\rho_s(0)\geqslant 0}$, i.e.~the staggered magnetization increases as a function of $U$.

If the interaction strength is much larger than the other parameters, i.e.~for 
${U\gg t_\uparrow,t_\downarrow, \Delta}$, the ground-state configuration has essentially one particle at each site 
(${\delta\rho_c \approx 0}$), and the low-energy degrees of freedom are spin $\frac{1}{2}$ operators ${\bf S}_i, i=1,\dots,L$.
Degenerate perturbation theory yields the Hamiltonian [\onlinecite{GMJ_16}]
\begin{align}
H&=J\sum_i\left(S_i^{(x)}S_{i+1}^{(x)}+S_i^{(y)}S_{i+1}^{(y)}+\gamma S_i^{(z)}S_{i+1}^{(z)}\right) \nonumber \\
 &-h\sum_i(-1)^iS_i^{(z)},
\label{eq:spinham}
\end{align}
with parameters
\begin{align}
J=\frac{4Ut_\uparrow t_\downarrow}{U^2-\Delta^2}\,,\quad 
\gamma=\frac{t_\uparrow^2+t_\downarrow^2}{2t_\uparrow t_\downarrow}\,,\quad
h=\frac{2\Delta(t_\uparrow^2-t_\downarrow^2)}{U^2-\Delta^2}\,.
\label{eq:parameters}
\end{align}
For ${t_\uparrow = t_\downarrow}$ we recover the isotropic Heisenberg chain, where the only effect of $\Delta$ is a slight enhancement of the exchange constant $J$. For ${t_\uparrow> t_\downarrow}$ and ${\Delta=0}$ we obtain an $XXZ$ chain with anisotropy parameter ${\gamma>1}$, i.e., a uniaxial antiferromagnet [\onlinecite{Kocharian_94, Fath95}]. For ${t_\uparrow> t_\downarrow}$ and ${\Delta>0}$ the $z$-components of the spins are coupled to a staggered longitudinal field of strength $h$. It is interesting to note that this Hamiltonian has been proposed in the context of quasi-one-dimensional easy-axis spin $\frac{1}{2}$ antiferromagnets (such as CsCoCl$_3$), where the staggered field arises in a mean-field treatment of interchain coupling [\onlinecite{Shiba_80}]. 

The Hamiltonian \eqref{eq:spinham} has been studied with different techniques, but mostly for isotropic exchange (${\gamma=0}$) or for the easy-plane case (${\gamma<1}$), where the interaction and the magnetic field compete [\onlinecite{Affleck_99, Lou_05, Mahdavifar_07, Paul_14}]. In the present case (${\gamma>1}$) the staggered field and the interaction reinforce each other in producing long-range antiferromagnetic order, with up-spins predominantly on even sites and down-spins mostly on odd sites.

\subsection{Ionic Hubbard model: $t_\uparrow=t_\downarrow$}
\label{sec:IH}
For spin-independent hopping (${t_\uparrow=t_\downarrow}$) the Hamiltonian
\eqref{eq:ham1} embodies the one-dimensional ionic Hubbard model, which has been studied intensively during recent decades 
[\onlinecite{Resta_Sorella_95, Ortiz_96, fab, tosatti_00, tor, Takada_Kido_01, Lou_03, brune, Manmana_04, Zhang_04, Solyom_06, Tincani_09, Hafez_10,Uhrig_2014}].
Early studies found two phases (at half filling), a band insulator for ${U\ll\Delta}$ and a Mott insulator for ${U\gg\Delta}$,
with a single quantum phase transition as a function of $U/\Delta$
[\onlinecite{Resta_Sorella_95,Ortiz_96,tosatti_00}]. 
The interest increased substantially when a field-theoretic treatment of the ionic Hubbard chain
came up with two quantum critical points $U_{c1}$ and $U_{c2}$ [\onlinecite{fab}]. The system was found to be a band insulator (with finite and almost equal spin and charge gaps) for ${U < U_{c1}}$ and a Mott insulator (with gapped charge and gapless spin excitations) for ${U > U_{c2}}$, as expected, but a new phase was found to sneak in for ${U_{c1}<U < U_{c2}}$. At ${U=U_{c1}}$ the charge gap vanishes and one finds metallic
behavior, the intermediate phase is a spontaneously dimerized insulator (with finite spin and
charge gaps), and at ${U=U_{c2}}$ the spin degrees of freedom exhibit a Kosterlitz-Thouless transition [\onlinecite{fab}]. Subsequent numerical studies
[\onlinecite{tor,brune,Manmana_04,Tincani_09}] clarified details of
these phases and unambiguously confirmed the scenario of two quantum phase
transitions [\onlinecite{Manmana_04,Tincani_09}]. 

The ionic Hubbard chain has been generalized by adding next-nearest-neighbor hopping [\onlinecite{Japaridze_07}] and by enlarging the unit cell to model for instance $MMX$ chains, where $M$ stands for metal atoms and $X$ for halogens  
[\onlinecite{Torio_06}]. In both cases rich phase diagrams have been found, where the
band- to Mott-insulator transition goes through a sequence of
unconventional insulating and/or metallic phases.
In higher dimensions the phase diagram is more controversial. The ionic Hubbard model for ${d>1}$ at half filling
has been studied by various methods, such as dynamical mean-field theory (DMFT)
[\onlinecite{Jabben_04, Randeria_06, Craco_07, Byczuk_09, Wang_14, Randeria_14, Krishnamurthy_15}],
quantum Monte Carlo [\onlinecite{Scalettar_07a, Scalettar_07b}],
cluster DMFT [\onlinecite{Kancharla_07}], a variational cluster approach [\onlinecite{Chen_10}] and the coherent potential approximation [\onlinecite{Hoang_10}]. Intermediate phases are routinely found, but depending on applied constraints and computational details different types of order emerge, from metallic 
[\onlinecite{Randeria_06,Craco_07,Scalettar_07a,Scalettar_07b,Hoang_10}], to half-metallic and antiferromagnetic  [\onlinecite{Randeria_14,Krishnamurthy_15}], as well as insulating and antiferromagnetic [\onlinecite{Kancharla_07,Byczuk_09,Chen_10}]. 

\subsection{Mass-imbalanced Hubbard model: $\Delta=0$}
In the absence of an alternating potential (${\Delta=0}$) the Hamiltonian \eqref{eq:ham1} represents the
mass-imbalanced Hubbard model, which was
introduced in the early nineties [\onlinecite{Brandt91}] to interpolate
between the standard Hubbard model (${t_{\uparrow}=t_{\downarrow}}$) and the Falicov-Kimball model 
(${t_\downarrow=0}$) [\onlinecite{Falicov_69}]. 

The one-dimensional mass-imbalanced Hubbard model is well understood at half filling in the large $U$ limit where it becomes equivalent to the easy-axis Heisenberg antiferromagnet, with anisotropy parameter $\gamma$ as in Eq.~\eqref{eq:parameters}. It is generally accepted [\onlinecite{Giamarchi_04, Mikeska_04}], that in this case the ground state has long-range antiferromagnetic order.
Moreover, the excitations are gapped. The situation is less clear for small and moderate values of $U$. Is there a (Kosterlitz-Thouless) transition to a gapped phase for an infinitesimal anisotropy [\onlinecite{Fath95}] or is the gapless phase extended over a finite region away from the line ${t_\uparrow=t_\downarrow}$ [\onlinecite{Chan_08}]?  
Away from half filling a lot of effort has been spent on the problem of phase separation, using rigorous techniques [\onlinecite{Ueltschi04}], bosonization [\onlinecite{Wang-Chen-Gu-07}], weak- and strong-coupling expansions [\onlinecite{Barbiero_10}], or numerical methods [\onlinecite{Silva-Valencia07, Gu-Fan-Lin-07,Farkasovsky08}]. 
Several studies were also made for higher dimensions, for instance on the Mott transition [\onlinecite{Winograd_11, Dao_12}] and on magnetic correlations [\onlinecite{Sotnikov_12, Liu_15}].

For ${t_\downarrow=0}$ we recover the Hamiltonian of the Falicov-Kimball model [\onlinecite{Falicov_69, Gruber_96, Freericks_03}]. Its ground state at half filling has been shown to consist of the most homogeneous configuration of immobile particles, which occupy the sites of one of the two sublattices [\onlinecite{Kennedy_86, Brandt_86}]. For other densities spin-up and spin-down particles are segregated for large enough repulsion [\onlinecite{Brandt_91, Freericks_02}]. In one dimension a more subtle phase separation can occur already for weak repulsion [\onlinecite{Freericks_96}].

\section{Mean-field theory}\label{sec:mft}
In the rest of this paper we seek the ground state of the mass-imbalanced ionic Hubbard chain in mean-field approximation. We only consider the broken symmetries included in the Hamiltonian \eqref{eq:ham1}, which lead to a period-two density modulation and spin alternation. If we would admit arbitrary densities we would have to include more complicated spatial orderings as well as phase separation and even ferromagnetism [\onlinecite{Farkasovsky_12}], but we restrict ourselves to half filling.
\subsection{Mean-field Hamiltonian and its ground state}
We define the mean-field state as the ground state of the single-particle Hamiltonian
\begin{align}
H_{\mbox{\scriptsize mf}}=-\sum_{i\sigma}\left[t_\sigma (c_{i\sigma}^\dag c_{i+1\sigma}^{\phantom{}}
+\mathrm{H.c.})-\frac{\Delta_\sigma}{2}(-1)^in_{i\sigma}\right]\!\!,
\end{align}
where $\Delta_\uparrow$ and $\Delta_\downarrow$ are variational parameters. 
This mean-field Hamiltonian can also be written as
\begin{align}
H_{\mbox{\scriptsize mf}}=&-\sum_{i\sigma}t_\sigma (c_{i\sigma}^\dag c_{i+1\sigma}^{\phantom{}} +\mathrm{H.c.})& \nonumber \\
                          &-\frac{L}{4}\big[(\Delta_\uparrow+\Delta_\downarrow)\delta\rho_c
                            +(\Delta_\downarrow-\Delta_\uparrow)\delta\rho_s\big].
\end{align}
Therefore $\Delta_\uparrow+\Delta_\downarrow$ and $\Delta_\uparrow-\Delta_\downarrow$ can be interpreted as conjugate fields of the order parameters $\delta\rho_c$ and $\delta\rho_s$, respectively.

$H_{\mbox{\scriptsize mf}}$ is diagonalized by the Bogoliubov transformation \eqref{eq:bog}, but now the ionic potential strength $\Delta$ has to be replaced by $\Delta_\sigma$ in Eqs.~\eqref{eq:angles} and \eqref{eq:spec}, i.e.,
\begin{align}
\tan 2\varphi_{k\sigma}=\frac{2\varepsilon_{k\sigma}}{\Delta_\sigma},\qquad
\cos 2\varphi_{k\sigma}=\frac{\Delta_\sigma}{2E_{k\sigma}},
\label{eq:angles2}
\end{align}
where
\begin{align}
E_{k\sigma}=\sqrt{\varepsilon_{k\sigma}^2+(\Delta_\sigma/2)^2}\,.
\label{eq:spec2}
\end{align}
The transformed Hamiltonian is again given by Eq.~\eqref{eq:trham} and its ground state (for ${N=L}$ particles) is
\begin{align}
\vert\Psi_0\rangle&=\prod_{k\sigma}\alpha_{k\sigma}^\dag\vert 0\rangle \nonumber \\
                  &=\prod_{k\sigma}\left(\cos\varphi_{k\sigma}\, a_{k\sigma}^\dag -\sin\varphi_{k\sigma}\, b_{k\sigma}^\dag\right)\vert0\rangle.
\label{eq:grst}
\end{align}
The order parameters are also calculated in exactly the same way as in Sec.~\ref{sec:nonint}, giving
\begin{align}
\begin{split}
\delta\rho_c&=\,\frac{1}{4\pi}\sum_\sigma\frac{\Delta_\sigma\kappa_\sigma K(\kappa_\sigma)}{t_\sigma}, \\
\delta\rho_s&=-\frac{1}{4\pi}\sum_\sigma\sigma\frac{\Delta_\sigma\kappa_\sigma K(\kappa_\sigma)}{t_\sigma},
\end{split}
\label{eq:ops2}
\end{align}
where
\begin{align}
\kappa_\sigma=\left[1+\left(\frac{\Delta_\sigma}{4t_\sigma}\right)^2\right]^{-1/2}.
\end{align}

\subsection{Mean-field energy and gap equations}

It is straightforward to calculate the expectation value of the Hamiltonian \eqref{eq:ham1}
with respect to the mean-field state \eqref{eq:grst}. We find

\begin{align}
&\frac{E(\Delta_\uparrow,\Delta_\downarrow)}{L}
=\sum_\sigma\left[\frac{\Delta_\sigma(\Delta_\sigma-\Delta)}{8\pi t_\sigma}\kappa_\sigma K(\kappa_\sigma)-\frac{2t_\sigma}{\pi}\frac{E(\kappa_\sigma)}{\kappa_\sigma}\right] \nonumber \\
&\qquad\quad+\frac{U}{4}\left[1+\frac{\Delta_\uparrow}{2\pi t_\uparrow}\kappa_\uparrow K(\kappa_\uparrow)
\frac{\Delta_\downarrow}{2\pi t_\downarrow}\kappa_\downarrow K(\kappa_\downarrow)\right].
\end{align}
To calculate derivatives, we use the following relations
\begin{align}
\begin{split}
\frac{d}{d\kappa_\sigma}\left[\frac{1}{\kappa_\sigma}E(\kappa_\sigma)\right]&=-\frac{1}{\kappa_\sigma^2}K(\kappa_\sigma), \\
\frac{d}{d\kappa_\sigma}\left[\kappa_\sigma K(\kappa_\sigma)\right]&=\frac{1}{1-\kappa_\sigma^2}E(\kappa_\sigma), \\
\frac{d}{d\Delta_\sigma}\left[\Delta_\sigma\kappa_\sigma K(\kappa_\sigma)\right]
&=\kappa_\sigma\left[K(\kappa_\sigma)-E(\kappa_\sigma)\right],
\end{split}
\label{eq:elliptic}
\end{align}
and obtain
\begin{align}
\begin{split}
\frac{1}{L}\frac{\partial E}{\partial\Delta_\uparrow}=A_\uparrow
\left[\Delta_\uparrow-\Delta+\frac{U}{2\pi t_\downarrow}\Delta_\downarrow\kappa_\downarrow K(\kappa_\downarrow)\right], \\
\frac{1}{L}\frac{\partial E}{\partial\Delta_\downarrow}=A_\downarrow
\left[\Delta_\downarrow-\Delta+\frac{U}{2\pi t_\uparrow}\Delta_\uparrow\kappa_\uparrow K(\kappa_\uparrow)\right],
\end{split}
\end{align}
with nonnegative coefficients
\begin{align}
A_\sigma=\frac{\kappa_\sigma}{8\pi t_\sigma}\left[K(\kappa_\sigma)-E(\kappa_\sigma)\right]\,\geqslant\,0\,.
\label{eq:coeff}
\end{align}

At extrema of the energy the first derivatives vanish, and we get the gap equations
\begin{align}
\begin{split}
\Delta_\uparrow=\Delta-\frac{U}{2\pi t_\downarrow}\Delta_\downarrow\kappa_\downarrow K(\kappa_\downarrow), \\
\Delta_\downarrow=\Delta-\frac{U}{2\pi t_\uparrow}\Delta_\uparrow\kappa_\uparrow K(\kappa_\uparrow).
\end{split}
\label{eq:gap}
\end{align}
They admit two different types of solutions, one where ${0<\Delta_\sigma<\Delta}$, and a second-one where the signs of $\Delta_\uparrow$ and $\Delta_\downarrow$ are different. These two possibilities correspond to different relative values of the order parameters \eqref{eq:ops2}.
If both gap parameters are positive, charge modulations dominate,
${\vert\delta\rho_c\vert>\vert\delta\rho_s\vert}$, but if the gap parameters have different signs, the alternating spin is dominant, 
${\vert\delta\rho_s\vert>\vert\delta\rho_c\vert}$. In general the gap equations have to be solved numerically, but for small on-site interaction we can use an expansion in $U$. To first order in $U$ we can obtain the mean-field corrections to the gap parameters by putting ${\Delta_\sigma=\Delta}$ on the r.h.s. of Eq.~\eqref{eq:gap}. Inserting these expressions into Eq.~\eqref{eq:ops2} we obtain the perturbative expressions \eqref{eq:pert}, which implies that mean-field theory is exact to leading order in $U$.

\section{Phases and phase transitions}\label{sec:pt}
We have solved numerically the gap equations \eqref{eq:gap} to determine the various phases and phase transitions of the model. The solution is not always unique. In such a case one can compare the energies to find out which solution corresponds to the minimum. Another useful criterion is local stability, which requires the second derivatives of the energy to be positive definite. 
For values $\Delta_\uparrow,\Delta_\downarrow$ satisfying the gap equations we obtain
\begin{align}
\frac{1}{L}\frac{\partial^2 E}{\partial\Delta_\sigma^2}=A_\sigma,\qquad
\frac{1}{L}\frac{\partial^2 E}{\partial\Delta_\uparrow\partial\Delta_\downarrow}=4UA_\uparrow A_\downarrow,
\label{eq:dm}
\end{align}
where the coefficients $A_\sigma$ are defined by Eq.~\eqref{eq:coeff}.
The eigenvalues of the ``dynamical matrix'' $\frac{1}{L}\left(\frac{\partial^2E}{\partial\Delta_\sigma\partial\Delta_{\sigma'}}\right)$ are
\begin{align}
\lambda_{\pm}=\frac{A_\uparrow+A_\downarrow\pm\sqrt{(A_\uparrow-A_\downarrow)^2+(8UA_\uparrow A_\downarrow)^2}}{2}\,.
\label{eq:eigenvalues}
\end{align}
At a local minimum both eigenvalues have to be positive. Solutions that do not satisfy this condition have to be ruled out as unstable. 

In some limiting cases analytical solutions of the gap equations can also be given. This will be particularly useful for discussing the behavior close to the transition.
In the following we choose $t_\uparrow=1$.
\subsection{From charge to spin order}
\label{sec:transition}
\begin{figure}
  \centering
  \includegraphics[width=\columnwidth]{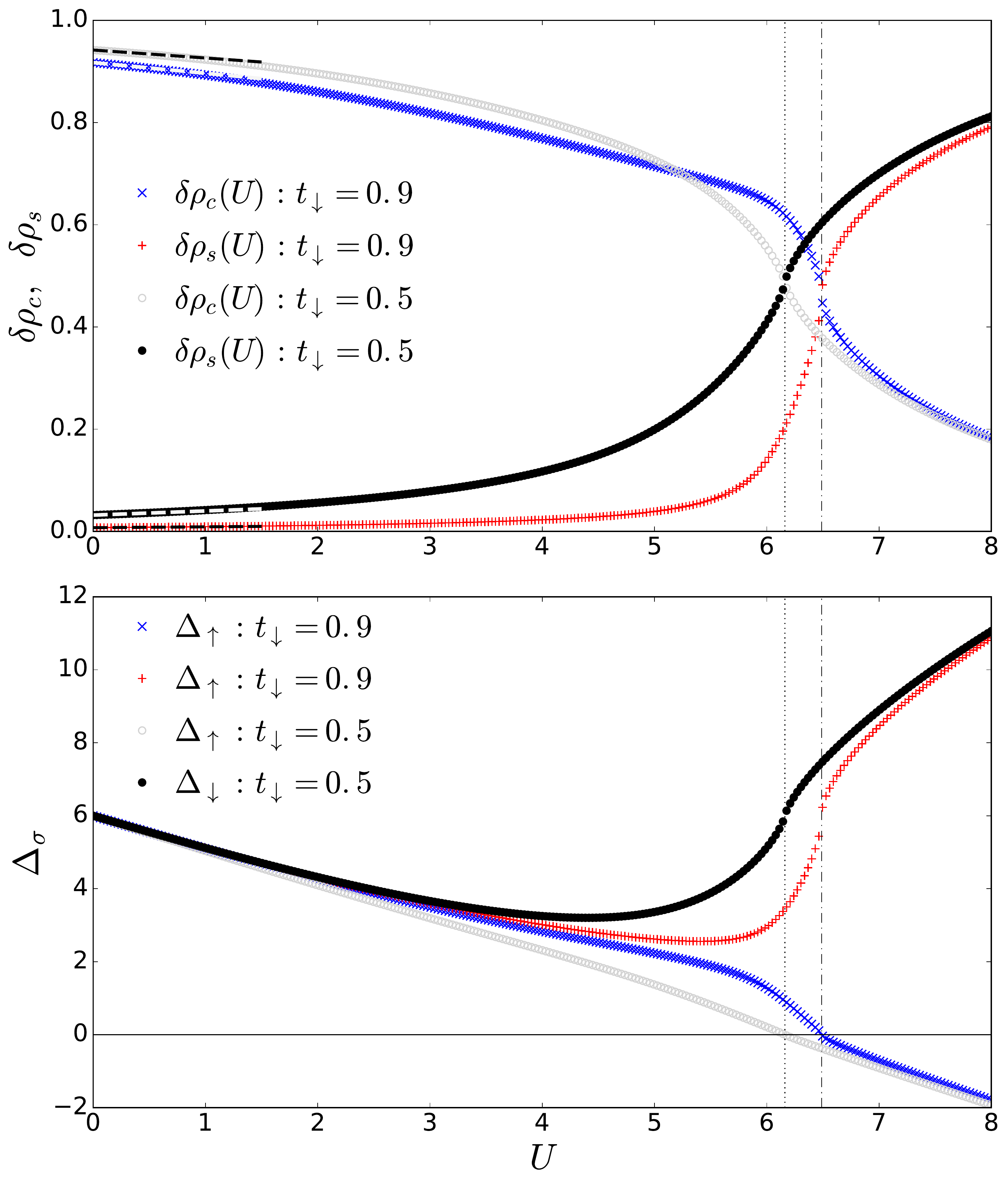}
  \caption{Order parameter (top) and gap parameters (bottom) for ${\Delta=6}$. For small $U$ the asymptotic solution \eqref{eq:pert} is indicated by dashed lines. Phase transition points ($U_c$) for ${t_\downarrow = 0.9}$ and ${t_\downarrow = 0.5}$ are indicated by dash-dot and dotted lines, respectively.}
  \label{fig:opsgaps2}
\end{figure}

\begin{figure}
  \centering
   \includegraphics[width=\columnwidth]{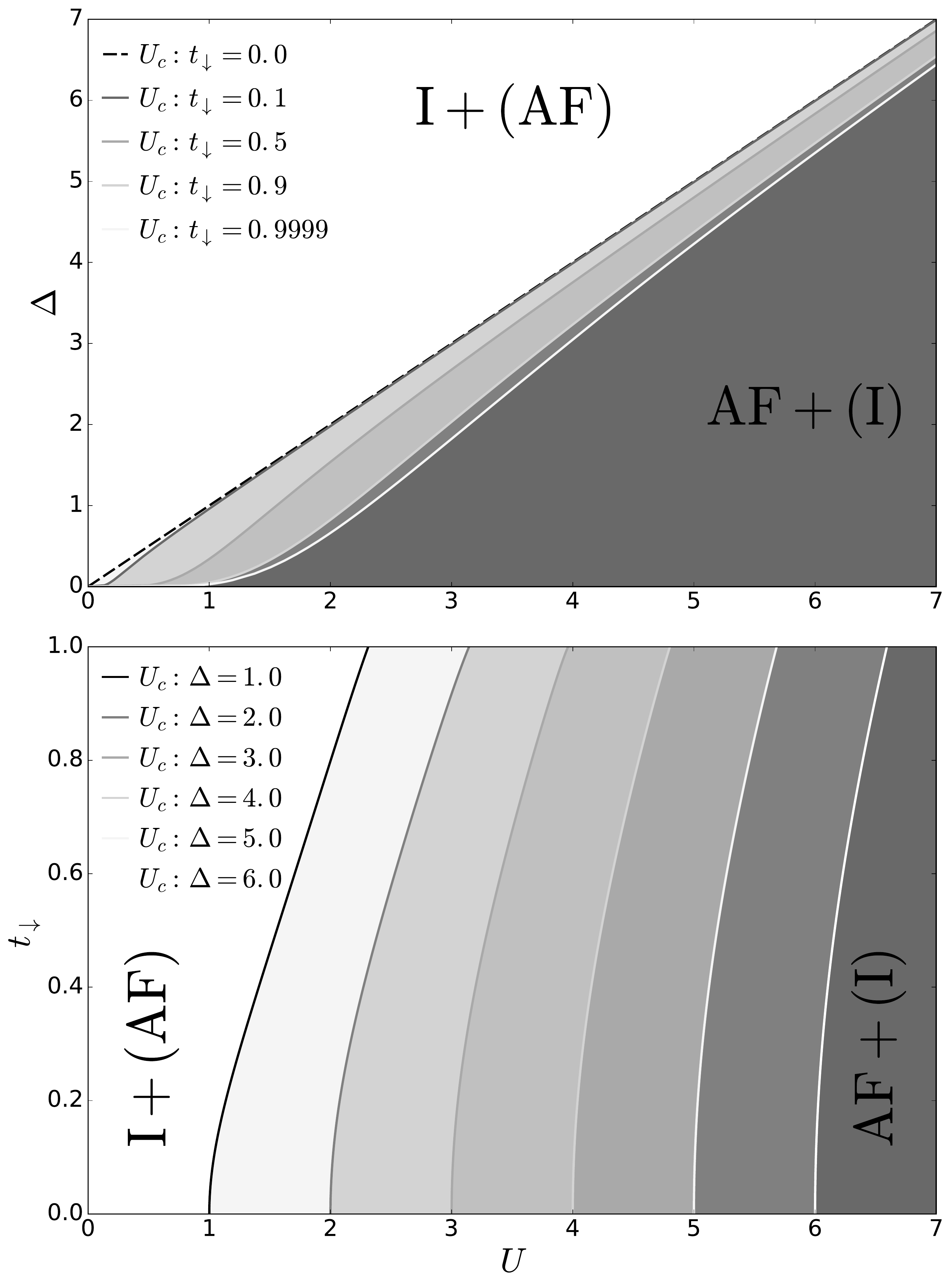}
   \caption{Mean-field phase diagrams, in the $U$-$\Delta$ plane for various anisotropies (top) and in the $U$-$t_\downarrow$ plane for various values of $\Delta$ (bottom). Long-range order exists for both the density modulation (ionic phase I) and spin alternation (AF) throughout the entire diagram. The notation I+(AF) stands for a dominant ionic phase (${\delta\rho_c>\delta\rho_s}$), while AF+(I) means that antiferromagnetism is stronger (${\delta\rho_s>\delta\rho_c}$).}
  \label{fig:pd1}
\end{figure}

Figure~\ref{fig:opsgaps2} shows the order parameters $\delta\rho_c$, $\delta\rho_s$ and the gap parameters 
$\Delta_\uparrow$, $\Delta_\downarrow$ for a potential strength ${\Delta=6}$ and two different anisotropies (${t_\downarrow=0.5}$ and $0.9$).  At ${U=U_c}$ there is a weakly first-order transition from dominant charge order to prevailing antiferromagnetism. The steps of the order parameters at $U_c$ decrease with decreasing $t_\downarrow$ and finally disappear as ${t_\downarrow\rightarrow 0}$.
The results for other values of $\Delta$ are qualitatively similar and lead to the phase diagram of 
Fig.~\ref{fig:pd1}. The transition line separating the ionic and antiferromagnetic phases depends only weakly on the mass imbalance $t_\downarrow/t_\uparrow$, especially for not too small values of $U$. For ${U\rightarrow\infty}$ the transition occurs at ${U=\Delta}$, as will be shown analytically below.

The barely visible discontinuities at $U_c$ (see Fig.~\ref{fig:opsgaps2}) are linked to the fact that the gap parameter $\Delta_\uparrow$ changes sign and, instead of passing smoothly through zero, jumps from a small positive value to a small negative value. 
A state with ${\Delta_\uparrow=0}$ would be metallic for the particles with up-spins and semiconducting for the others. From the point of view of symmetry, the full translational invariance of the mean-field Hamiltonian would be restored for one spin component. A solution of Eq.~\eqref{eq:gap} with ${\Delta_\uparrow=0}$ indeed exists, with ${\Delta_\downarrow=\Delta}$ (for a particular value $U^*$). However, when this point is approached, the quantity $A_\uparrow$, defined by Eq.~\eqref{eq:coeff}, diverges, while $A_\downarrow$ remains finite. Therefore the eigenvalue $\lambda_-$ of Eq.~\eqref{eq:eigenvalues} is negative and the solution is unstable. This result is a nice example for the resistance of many-body systems against the restoration of an explicitly broken symmetry.

\begin{figure}
  \centering
   \includegraphics[width=\columnwidth]{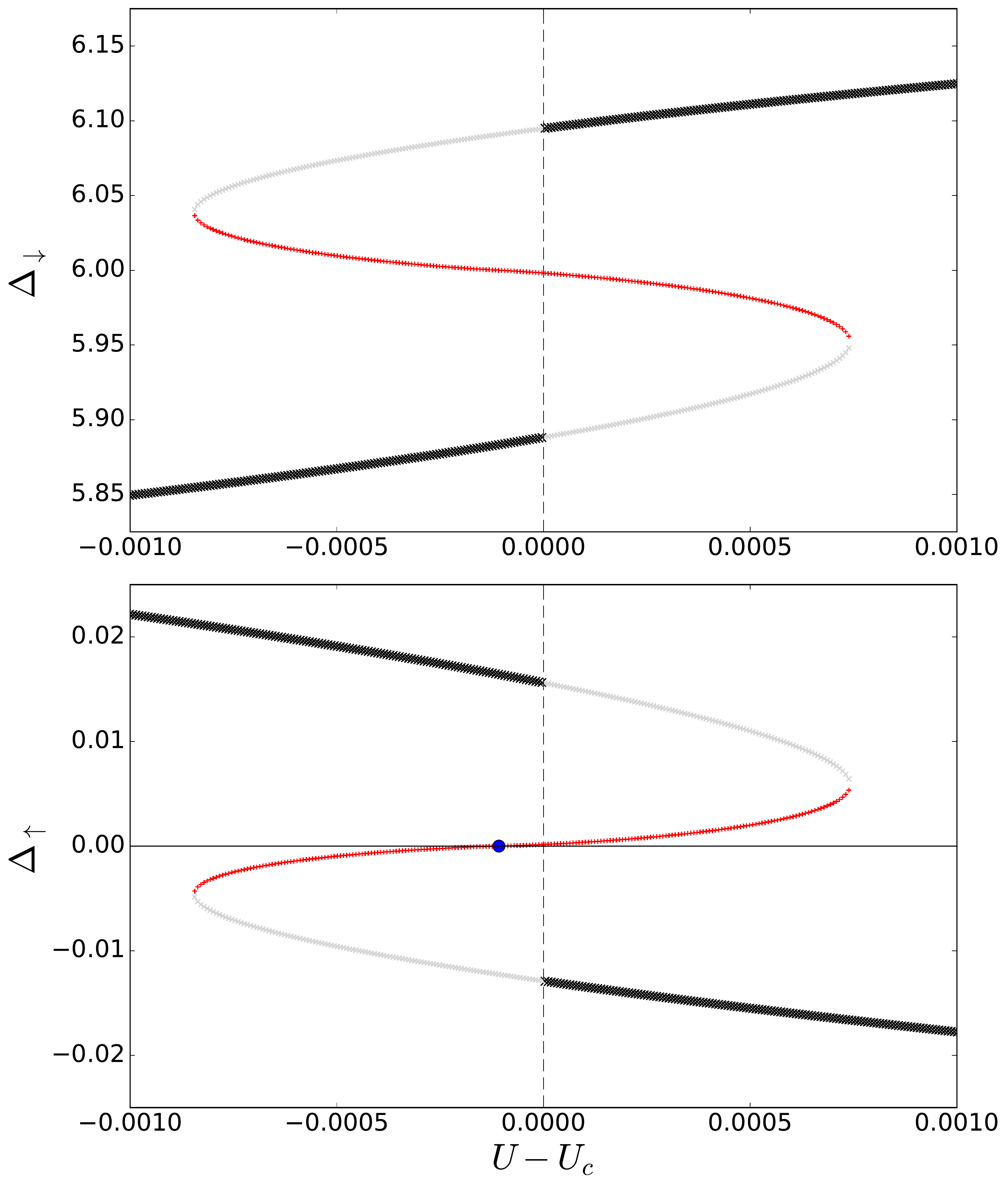}
   \caption{Solution of the gap equation close to $U_c$ for ${t_\uparrow=1}$ and ${t_\downarrow=0.9}$. The black segments stand for the most stable solution, while the middle segment (red) represents an unstable solution. A zero-crossing point of the unstable solution ($U^{*}$) is indicated by a blue dot.}
  \label{fig:critical}
\end{figure}

To see how criticality is avoided, we show the solution of the gap equations close to $U_c$ in Fig.~\ref{fig:critical}. The upper and lower lines of the $z$- and $s$-shaped curves correspond to locally stable solutions (${\lambda_->0}$), while the middle section represents an unstable solution (${\lambda_-<0}$). In fact, we can easily show that the stability limit 
(${\lambda_-=0}$) coincides with the point where $d\Delta_\sigma/dU$ diverges. From Eq.~\eqref{eq:gap} we obtain
\begin{align}
\frac{d\Delta_\sigma}{dU}=\frac{1}{1-16U^2A_\uparrow A_\downarrow}
\left[\frac{\Delta_\sigma-\Delta}{U}-4A_{\bar{\sigma}}(\Delta_{\bar{\sigma}}-\Delta)\right].
\label{eq:derivative}
\end{align}
In view of Eq.~\eqref{eq:eigenvalues} the prefactor diverges when ${\lambda_-\rightarrow 0}$.
The value $U_c$ is calculated by comparing the energies of the two locally stable branches; the upper line has a smaller energy than the lower one for ${U<U_c}$ and the lower one is preferable for ${U>U_c}$. 

We estimate $U_c$ for large values of $\Delta$ where the hysteresis is very small (it vanishes for ${\Delta\rightarrow\infty}$) and ${U_c\approx U^*}$. At the unstable critical point (with ${\Delta_\uparrow=0}$ and ${\Delta_\downarrow=\Delta}$) the gap equations boil down to
\begin{align}
1=\frac{U^*}{2\pi t_\downarrow}\kappa_\downarrow^0K(\kappa_\downarrow^0), 
\label{eq:gap3}
\end{align}
where $\kappa_\downarrow^0=[1+(\Delta/4t_\downarrow)^2]^{-\frac{1}{2}}$.
For large values of $\Delta$ the parameter $\kappa_\downarrow^0$ is very small, $\kappa_\downarrow^0\approx 4t_\downarrow/\Delta\ll 1$, and we find
\begin{align}
U_c\approx U^*\approx \Delta\left[1+\left(\frac{2t_\downarrow}{\Delta}\right)^2\right] 
\label{eq:uc}
\end{align}
and therefore $U_c$ indeed tends to $\Delta$ for ${\Delta\rightarrow\infty}$.

To estimate the discontinuity in the gap parameters at $U_c$ we determine the three solutions of the gap equations at $U^*$, again in the large $\Delta$ limit, where $\Delta_\uparrow$ is very small and 
$\Delta_\downarrow$ is close to $\Delta$. To leading orders of $\Delta_\uparrow$ and $({\Delta_\downarrow-\Delta})$ the gap equations read
\begin{align}
\begin{split}
\Delta_\uparrow\approx&-\frac{8U^*t_\downarrow^2}{\Delta^3}\, (\Delta_\downarrow -\Delta), \\
\Delta_\downarrow-\Delta\approx&-\frac{U^*}{2\pi t_\uparrow}\Delta_\uparrow\log\frac{4t_\uparrow}{\vert\Delta_\uparrow\vert},
\end{split}
\end{align}
where Eq.~\eqref{eq:gap3} has been used. Besides the unstable solution there exist two other solutions, corresponding to the stable branches, namely
\begin{align}
\Delta_\uparrow &\approx\pm 4t_\uparrow \exp\left(-\frac{\pi\Delta^3t_\uparrow}{4(U^*)^2t_\downarrow}\right) \nonumber \\
               &\approx 4t_\uparrow \exp\left({-\frac{\pi\Delta t_\uparrow}{4t_\downarrow^2}}\right).
\end{align}
The jump ${2\vert\Delta_\uparrow\vert}$ vanishes exponentially both for ${\Delta\rightarrow\infty}$ and for ${t_\downarrow\rightarrow 0}$. The same exponential behavior is found for $\Delta_\downarrow$.

\subsection{Fidelity susceptibility}
In addition to order parameters an interesting quantity characterizing the ground state is the fidelity susceptibility, which has been used successfully for quantum phase transitions [\onlinecite{Zanardi_06, Gu_10}], and also for the crossover from Bose-Einstein condensation of tightly bound pairs to BCS superconductivity [\onlinecite{Khan_09}]. 
For a given Hamiltonian, which depends on some parameter, typically a dimensionless coupling constant, the fidelity is defined as the overlap of the ground states for two different values of this parameter. In the present case we choose $U$ (for ${t_\uparrow=1}$) as characteristic parameter and use the (normalized) mean-field ground state ${\vert\Psi(U)\rangle}$ to calculate the fidelity 
\begin{align}
F(U_1,U_2)=\langle\Psi(U_1)\vert\Psi(U_2)\rangle\,.
\end{align}  
The fidelity susceptibility is then defined as
\begin{align}
\chi_F(U):=-\frac{2}{L}\lim_{\delta U\rightarrow 0}
\frac{\log F(U,U+\delta U)}{(\delta U)^2}\,.
\end{align}
With our mean-field ground state \eqref{eq:grst} we obtain
\begin{align}
\chi_F(U)
=\frac{1}{L}\sum_{k\sigma}\left(\frac{d\varphi_{k\sigma}}{dU}\right)^2.
\label{eq:chi1}
\end{align}
The Bogoliubov angles $\varphi_{k\sigma}$ are linked to the gap parameters $\Delta_\sigma$ by Eq.~\eqref{eq:angles2}, i.e.
\begin{align}
\frac{d\varphi_{k\sigma}}{d\Delta_\sigma}=-\frac{\varepsilon_{k\sigma}}{4E_{k\sigma}^2},
\end{align}
and the derivatives of $\Delta_\sigma$ with respect to $U$ are given by Eq.~\eqref{eq:derivative}.
Working out the $k$-sum in Eq.~\eqref{eq:chi1} for ${L\rightarrow\infty}$ we finally get
\begin{align}
\chi_F(U)=\sum_\sigma \left(\frac{1}{16t_\sigma}\right)^2\frac{\kappa_\sigma^4}{\sqrt{1-\kappa_\sigma^2}}\left(\frac{d\Delta_\sigma}{dU}\right)^2.
\end{align}
This expression is valid both for $U<U_c$ and for $U>U_c$.
For a first-order transition at $U=U_c$, where the ground states above and below the transition point differ, the fidelity $F(U_c-\frac{1}{2}\delta U,U_c+\frac{1}{2}\delta U)$ is smaller than 1, even in the limit $\delta U\rightarrow 0$, and $\chi_F(U_c)=\infty$. We expect therefore a sharp line emerging from a smooth background, in contrast to the case of a second-order transition, where $\chi_F$ diverges when a critical point is approached.

\begin{figure}
  \centering
   \includegraphics[width=\columnwidth]{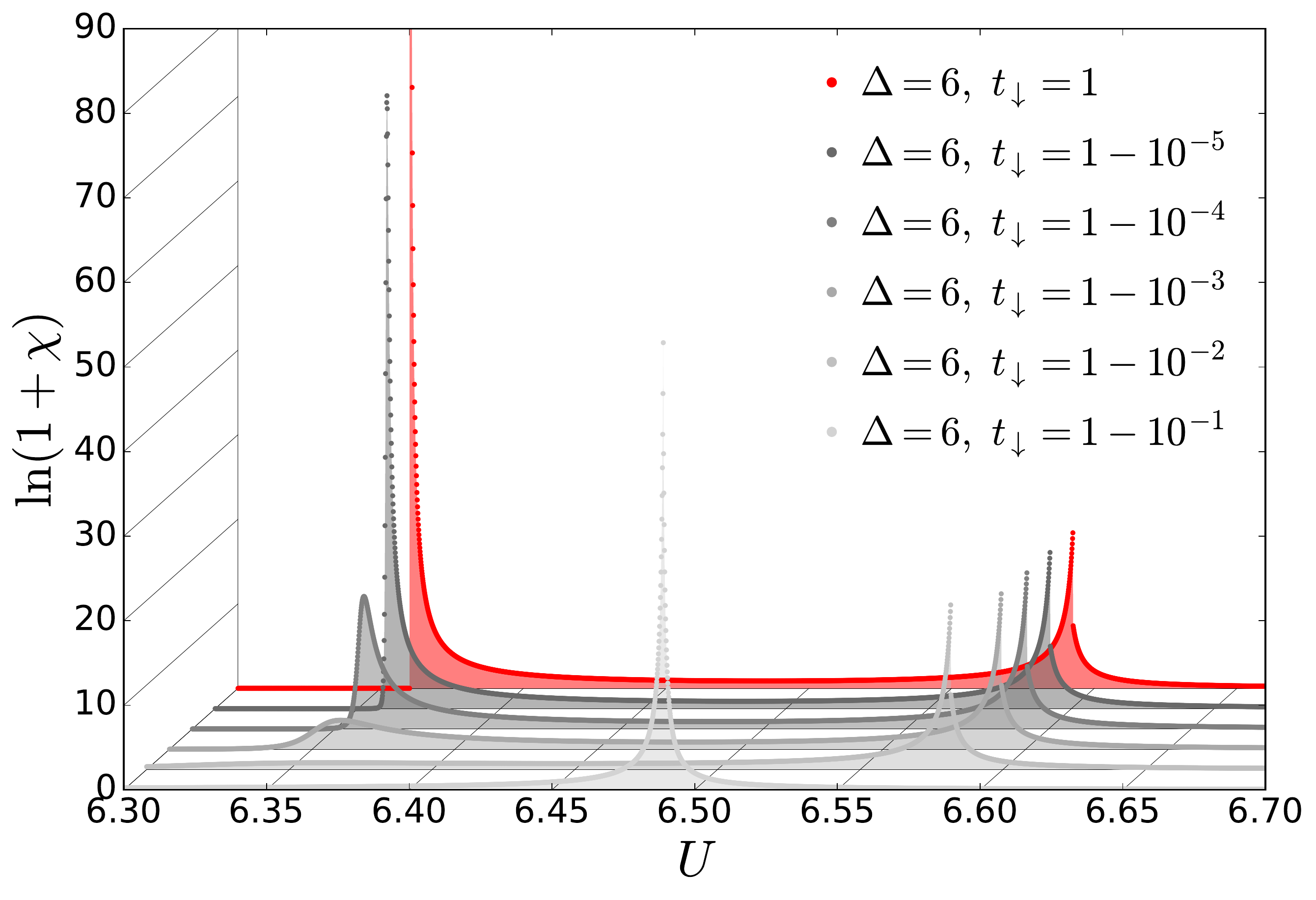}
   \caption{Fidelity susceptibility in the vicinity of $U_c$ for ${\Delta=6}$ and various hopping anisotropies.}
  \label{fig:fid1}
\end{figure}

Figure~\ref{fig:fid1} shows the fidelity susceptibility for ${\Delta=6}$ and various values of $t_\downarrow$. For ${t_\downarrow<0.99}$ a single peak is found, and it is located at $U_c$. We notice that this line is not simply an infinitely sharp peak as in an ordinary first-order transition, but $\chi_F$ is strongly enhanced upon approaching $U_c$, in agreement with the avoided criticality discussed above. Mathematically, this enhancement comes from the first factor of the r.h.s. of Eq.~\eqref{eq:derivative}, which can also be written as $A_\uparrow A_\downarrow/(\lambda_+\lambda_-)$. The eigenvalue $\lambda_-$ decreases when $U_c$ is approached and is very small at $U_c$. 
Surprisingly, a second peak appears if $t_\downarrow$ approaches 1. It has the form of a weak hump for ${t_\downarrow=0.999}$, which becomes sharper as $t_\downarrow$ approaches 1 and finally is singular at the isotropic point, ${t_\downarrow=1}$. We attribute the hump to a crossover due to a rapid change in gap parameters, whereas the singularity for ${t_\downarrow=1}$ marks a second-order phase transition, as will be discussed in more detail in the next section.

\subsection{Mean-field theory for the ionic Hubbard model}
\begin{figure}[t]
  \centering
   \includegraphics[width=\columnwidth]{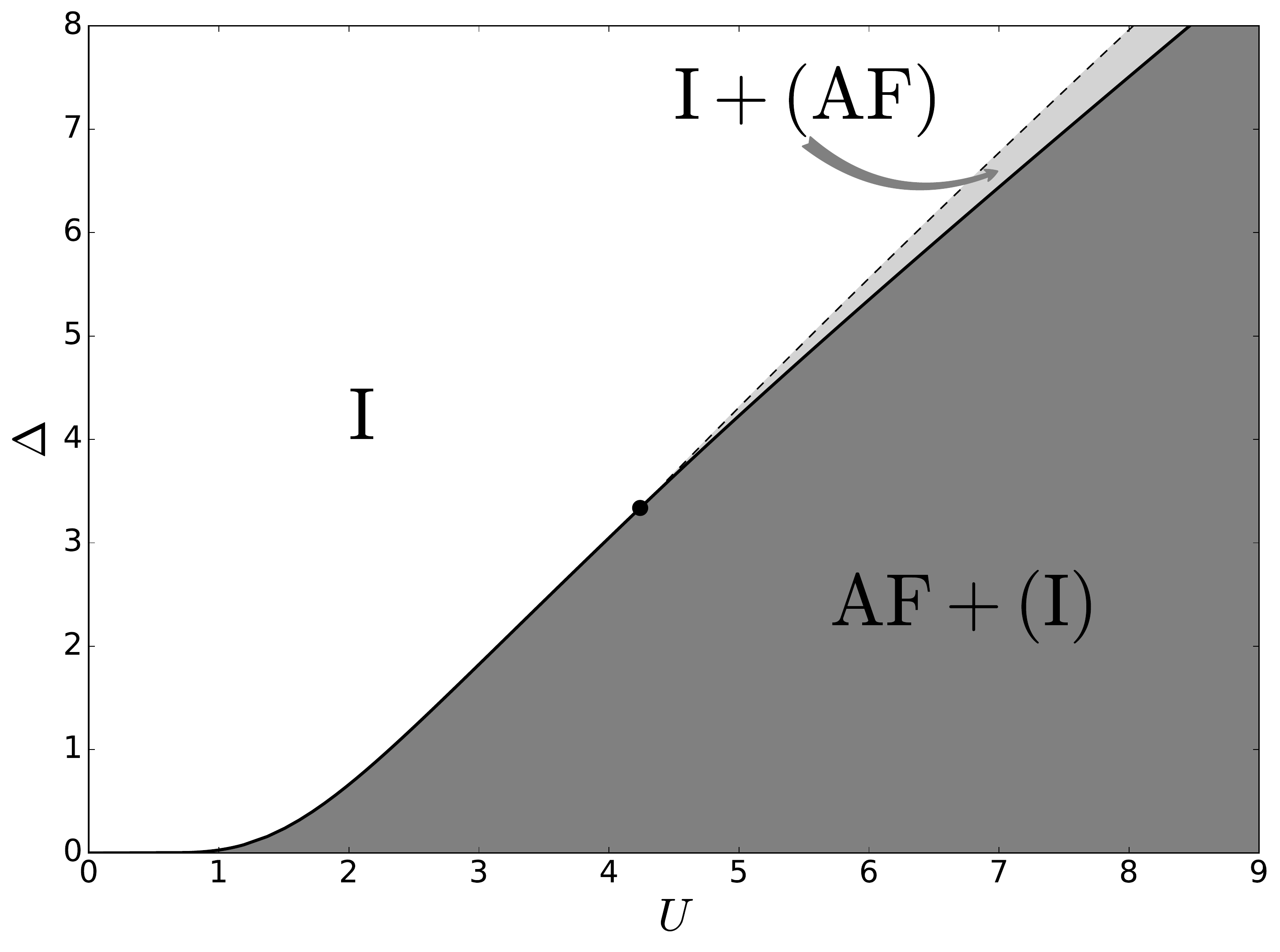}
   \caption{Mean-field phase diagram for the ionic Hubbard model (${t_\uparrow=t_\downarrow=1}$). }
  \label{fig:pd2}
\end{figure}

Figure~\ref{fig:pd2} shows the phase diagram for the ionic Hubbard chain. The main differences with respect to the mass-imbalanced case of Fig.~\ref{fig:pd1} are the existence of a pure ionic phase, where ${\delta\rho_s=0}$, and the appearance of the intermediate phase.
The bifurcation from a single transition line to two lines occurs at $\Delta\approx 3.3373088,\, U\approx 4.2398854$.

\begin{figure}[t]
  \centering
   \includegraphics[width=\columnwidth]{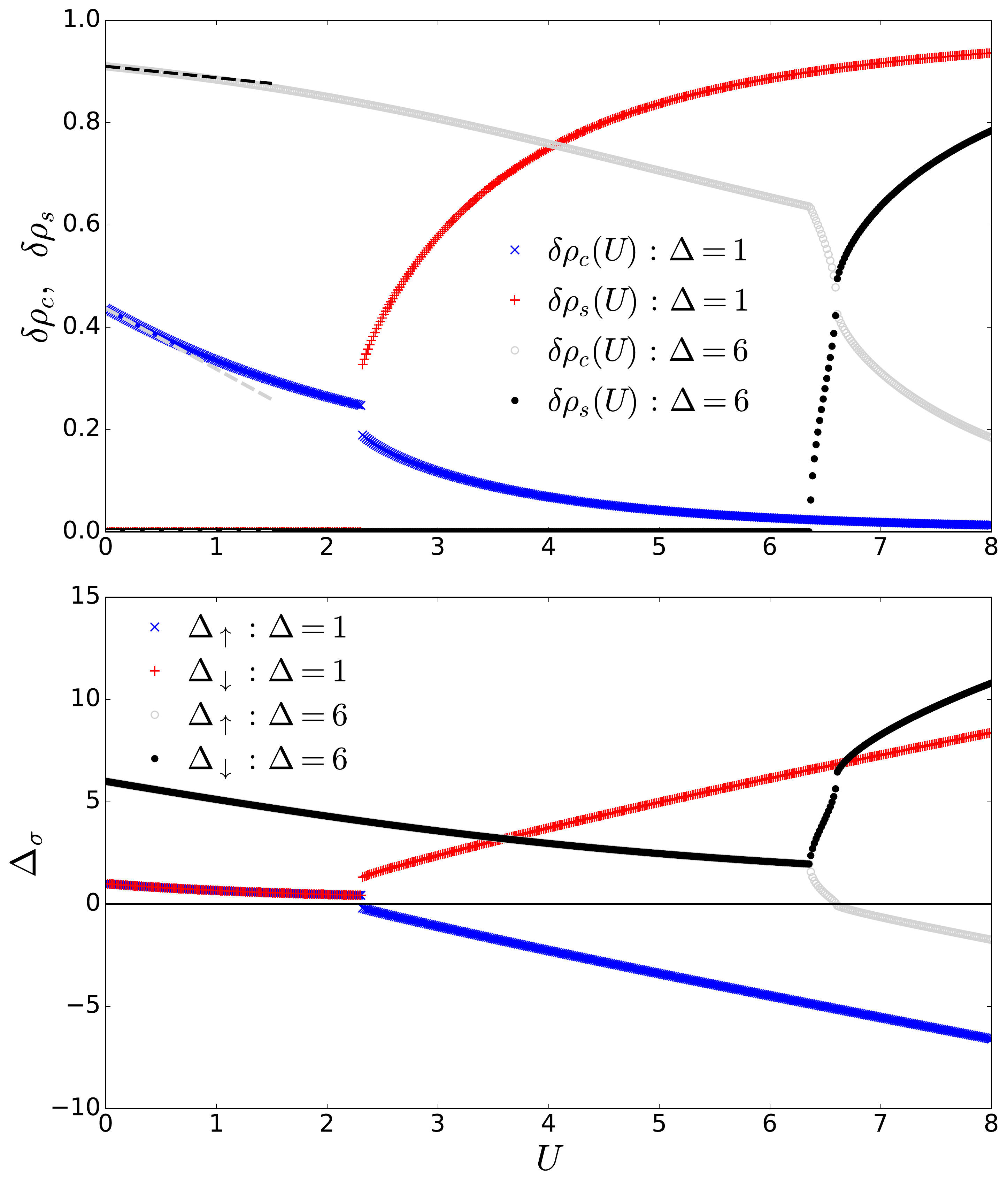}
   \caption{Order parameter (top) and gap parameters (bottom) for the ionic Hubbard model (${t_\uparrow=t_{\downarrow}=1}$) for several values of $\Delta$.}
  \label{fig:opsgaps1}
\end{figure}
We now turn our attention to the ionic Hubbard model (${t_\uparrow=t_\downarrow}$), where the spin SU(2) symmetry is not explicitly broken.
Figure~\ref{fig:opsgaps1} shows the results for the gaps $\Delta_\sigma$ and the order parameters $\delta\rho_c$, $\delta\rho_s$ as functions of $U$. For small $U$ the gap parameters are equal and, correspondingly, there is no antiferromagnetism. For large $U$ the gap parameters have different signs and therefore the antiferromagnetic order parameter $\delta\rho_s$ dominates. The nature of the transition between these two regimes depends on the strength of the ionic potential $\Delta$. For small $\Delta$ there is a single first-order transition, as in the mass-imbalanced case. For large $\Delta$ there are two transitions, first a continuous transition at $U_{c1}$ where the difference between gap parameters starts from zero and both of them remain first positive. In this intermediate phase the density modulation still dominates, but there is already antiferromagnetic order. A second transition occurs at a slightly larger value $U_{c2}$ and is of first order. For ${U>U_{c2}}$ antiferromagnetism dominates. It is worthwhile to mention that the two transitions at $U_{c1}$ and $U_{c2}$, respectively, coincide with the two singularities observed in the fidelity susceptibility for ${t_\uparrow=t_\downarrow}$ (see Fig.~\ref{fig:fid1}).

We focus first on the continuous transition at $U_{c1}$ and show that it is triggered by the softening of antiferromagnetic fluctuations. We start within the purely ionic phase, ${U<U_{c1}}$, where a single gap parameter $\Delta_\uparrow=\Delta_\downarrow=:\bar{\Delta}$ satisfies the equation
\begin{align}
\bar{\Delta}=\Delta-\frac{U}{2\pi t}\bar{\Delta}\bar{\kappa}K(\bar{\kappa}).
\label{eq:gap_av}
\end{align}
We now enhance $U$ by an infinitesimal amount $\delta U$ and assume the gap parameters to change accordingly,
\begin{align}
\Delta_\uparrow=\bar{\Delta}(1+\eta_\uparrow),\qquad \Delta_\downarrow=\bar{\Delta}(1+\eta_\downarrow).
\label{eq:parametrization}
\end{align}
Expanding the gap equations \eqref{eq:gap} up to first order in $\delta U$ and $\eta_\sigma$ and subtracting Eq.~\eqref{eq:gap_av} we get the linear system
\begin{align}
\begin{split}
\eta_\uparrow=-\frac{1}{2\pi t}\Big\{U\bar{\kappa}\left[K(\bar{\kappa})-E(\bar{\kappa})\right]\eta_\downarrow
+\bar{\kappa}K(\bar{\kappa})\delta U\Big\}, \\
\eta_\downarrow=-\frac{1}{2\pi t}\Big\{U\bar{\kappa}\left[K(\bar{\kappa})-E(\bar{\kappa})\right]\eta_\uparrow
+\bar{\kappa}K(\bar{\kappa})\delta U\Big\}.
\end{split}
\end{align}
Adding and subtracting these two equations we obtain
\begin{align}
\Big\{1+\frac{\bar{\kappa}U}{2\pi t}\left[K(\bar{\kappa})-E(\bar{\kappa})\right]\Big\}\left(\eta_\uparrow+\eta_\downarrow\right)
&=-\frac{\bar{\kappa}}{\pi t}K(\bar{\kappa})\, \delta U,\nonumber\\
\Big\{1-\frac{\bar{\kappa}U}{2\pi t}\left[K(\bar{\kappa})-E(\bar{\kappa})\right]\Big\}\left(\eta_\uparrow-\eta_\downarrow\right)&=0\,.
\end{align}
The first line implies that the sum of the gap parameters varies linearly with $\delta U$, while the second line implies ${\eta_\uparrow=\eta_\downarrow}$, {\it as long as the prefactor is finite}. This prefactor is proportional to the eigenvalue $\lambda_-$ and therefore a new solution ${\eta_\uparrow\neq\eta_\downarrow}$ appears at the particular value of $U$ where  ${\lambda_-=0}$, which we therefore identify with the critical point $U_{c1}$ for the transition to the intermediate phase. 

We can also use the parametrization \eqref{eq:parametrization} for calculating the eigenvectors of the dynamical matrix \eqref{eq:dm}. One readily finds ${\eta_\uparrow=\pm\eta_\downarrow}$ for $\lambda_\pm$. The eigenvector of $\lambda_-$ modifies primarily 
$\delta\rho_s$ and can therefore be associated with a magnetic excitation. This completes the soft-mode picture of the transition at the critical 
point $U_{c1}$, which is determined by the gap equation \eqref{eq:gap_av} (with $U=U_{c1}$) together with the soft-mode condition
\begin{align}
1=4U_{c1}\bar{A}=\frac{\bar{\kappa}U_{c1}}{2\pi t}[K(\bar{\kappa})-E(\bar{\kappa})].
\label{eq:sm}
\end{align}

\begin{figure}
  \centering
   \includegraphics[width=\columnwidth]{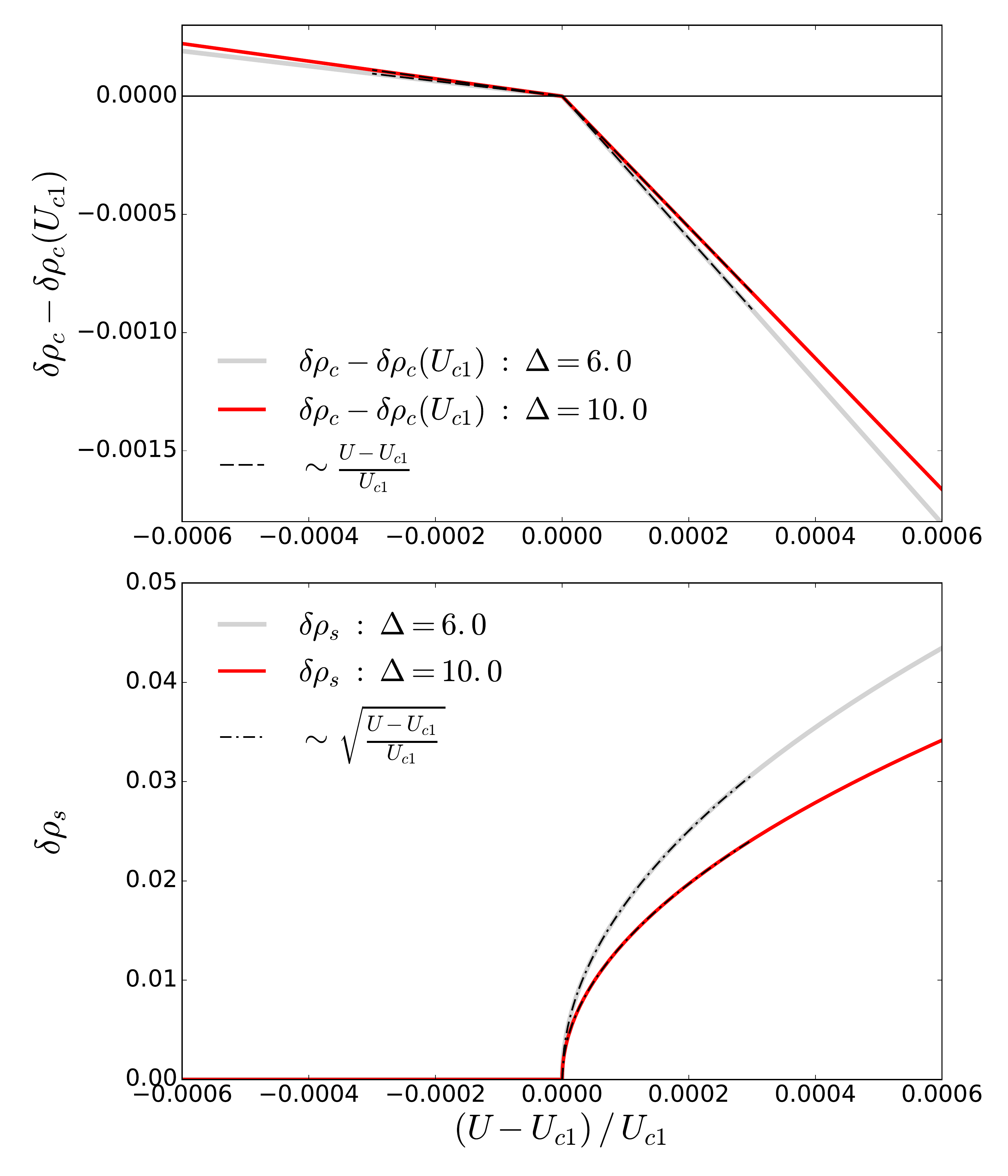}
   \caption{Critical behavior of the order parameters for the ionic Hubbard model at $U_{c1}$ for two different values of $\Delta$. Full lines are numerical data, dashed lines represent the analytical results of Eqs.~\eqref{eq:spin} and \eqref{eq:charge}.}
  \label{fig:uc1}
\end{figure}

When $U_{c1}$ is approached from below, the order parameters are
\begin{align}
&\delta\rho_s=0, \nonumber \\
&\frac{\delta\rho_c(U)-\delta\rho_c(U_{c1})}{\delta\rho_c(U_{c1})} \approx\frac{U_{c1}-U}{2U_{c1}}.
\label{eq:opbelow}
\end{align}
The treatment of the region above the critical point is slightly
more tricky because we have to use a higher-order expansion. The calculation, detailed in the Appendix, gives the following result for the critical region
\begin{align}
\begin{split}
&\delta\rho_s\sim\left(\frac{U-U_{c1}}{U_{c1}}\right)^{\frac{1}{2}}, \\
& \delta\rho_c(U)-\delta\rho_c(U_{c1})\sim\left(\frac{U-U_{c1}}{U_{c1}}\right),
\end{split}
\label{eq:opabove}
\end{align}
with proportionality factors given by Eqs. (\ref{eq:spin}) and (\ref{eq:charge}), respectively.
The behavior of Eqs. (\ref{eq:opbelow}) and (\ref{eq:opabove}) agrees perfectly well with
numerical calculations in the vicinity of $U_{c1}$, as shown in Fig. \ref{fig:uc1}.

In Fig.~\ref{fig:pd2} the intermediate phase appears as a tiny fjord, separating the dominating ionic and antiferromagnetic regions. To find out how this fjord widens further up we calculate the asymptotic behavior of the two transition lines. The transition at $U_{c2}$ is exactly of the same nature as the single transition at $U_c$ found in the case of unequal hopping parameters and therefore we can use the large $\Delta$ estimate of Eq.~\eqref{eq:uc} for $U_{c2}$.
The critical point $U_{c1}$ is determined by two equations, the gap equation \eqref{eq:gap_av} and the soft-mode condition \eqref{eq:sm}.
For large $\Delta$ $U_{c1}$ is also large and therefore the parameter $\bar{\kappa}$ has to be small. 
Expanding $\bar{\Delta}$ and the elliptic integrals in powers of $\bar{\kappa}$ we get
\begin{align}
\frac{\Delta}{8t}\sim\frac{1}{\bar{\kappa}^3}\left(1-\frac{1}{8}\bar{\kappa}^2\right)
\end{align}
together with
\begin{align}
\frac{U_{c1}}{8t}\sim \frac{1}{\bar{\kappa}^3}\left(1-\frac{3}{8}\bar{\kappa}^2\right)
\end{align}
and therefore
\begin{align}
\frac{U_{c1}}{\Delta}\sim1-\frac{1}{4}\bar{\kappa}^2\sim 1-\left(\frac{t}{\Delta}\right)^{\frac{2}{3}}.
\end{align} 
We conclude that the size of the intermediate region steadily grows and diverges for ${\Delta\rightarrow\infty}$, 
\begin{align}
U_{c2}-U_{c1}\sim(t^2\Delta)^{\frac{1}{3}}.
\end{align}

Coming back to the fidelity susceptibility (see Fig.~\ref{fig:fid1}) we recall that for the ionic Hubbard model (${t_\uparrow=t_\downarrow}$) we found two peaks, one at $U_{c2}$ with a similar shape as that found for the mass-imbalanced case and one at $U_{c1}$ with a very different form. When approaching $U_{c1}$ from above we find a divergence characteristic of a second-order phase transition. This is not seen when approaching $U_{c1}$ from below. We can readily understand this interesting result from Eq.~\eqref{eq:derivative}, because for ${U<U_{c1}}$
the singularity due to the first factor (triggered by the soft mode) is compensated by the second factor and we get
\begin{align}
\frac{d\bar{\Delta}}{dU}=(1+4U\bar{A})^{-1}\, \frac{\bar{\Delta}-\Delta}{U}\,.
\end{align}

Does the peculiar behavior of the fidelity susceptibility for ${U<U_{c1}}$  recur in other quantities, for instance in the staggered magnetic susceptibility? To calculate the response to a staggered magnetic field we add the perturbation
\begin{align}
H'=-\frac{h}{2}\sum_{i\sigma}(-1)^i\sigma n_{i\sigma}
\end{align}
to the Hamiltonian \eqref{eq:ham1}. Its expectation value with respect to the mean-field ground state is
\begin{align}
\langle H'\rangle=\frac{Lh}{8\pi t}\sum_\sigma\sigma\Delta_\sigma\kappa_\sigma K(\kappa_\sigma)
\end{align}
and the derivatives with respect to the gap parameters are
\begin{align}
\frac{\partial}{\partial\Delta_\sigma}\langle H'\rangle=Lh\sigma A_\sigma.
\end{align}
Therefore the gap equations read
\begin{align}
\begin{split}
\Delta_\uparrow=\Delta-h-\frac{U}{2\pi t}\Delta_\downarrow\kappa_\downarrow K(\kappa_\downarrow), \\
\Delta_\downarrow=\Delta+h-\frac{U}{2\pi t}\Delta_\uparrow\kappa_\uparrow K(\kappa_\uparrow).
\end{split}
\label{eq:gap_field}
\end{align}
We seek a solution as in Eq.~\eqref{eq:parametrization}
where $\bar{\Delta}$ is the gap parameter for ${h=0}$. For an infinitesimal value of $h$ we find
\begin{align}
\eta_\sigma=-\frac{\sigma h}{\bar{\Delta}(1-4U\bar{\Delta})}\,.
\end{align}
The staggered magnetic response is measured by the susceptibility
\begin{align}
\chi_{\mbox{\scriptsize st}}:=&\, \lim_{h\to 0}\, \frac{1}{h}\left[\delta\rho_s(h)-\delta\rho_s(0)\right] \nonumber \\
=&-2\bar{A}\bar{\Delta}\, \lim_{h\to 0}\, \frac{\eta_\uparrow-\eta_\downarrow}{h}=\frac{4\bar{A}^2}{\lambda_-}\,.
\end{align}
The eigenvalue $\lambda_-=\bar{A}(1-4U\bar{A})$ vanishes as ${(U_{c1}-U)}$ because $\bar{A}$ is finite and continuous close to $U_{c1}$. Therefore the staggered susceptibility diverges as ${(U_{c1}-U)^{-1}}$ and does not exhibit any anomalous behavior.

At this point we have to compare the mean-field phase diagram of Fig.~\ref{fig:pd2} with that obtained using more elaborate methods (mentioned in Sec.~\ref{sec:IH}). There is qualitative agreement for the ionic phase, the location of the transition line and even the existence of an intermediate phase. However, the nature of both antiferromagnetic and intermediate phases differs. While we found long-range magnetic order in both phases, only quasi-long-range order is obtained by essentially exact treatments for the antiferromagnetic phase. This can be readily understood in terms of quantum fluctuations, which are neglected in mean-field theory. However, in the intermediate phase the ordering found both in numerical simulations and in analytical treatments for small $U$ is completely different from our mean-field result; this disparity cannot be simply attributed to order parameter fluctuations.

\subsection{The ionic Falicov-Kimball model: $t_\downarrow=0$}
For ${t_\downarrow\rightarrow 0}$ the gap equations are simplified considerably. With the limiting behavior
\begin{align}
\frac{\Delta_\downarrow\kappa_\downarrow K(\kappa_\downarrow)}{t_\downarrow}\rightarrow 2\pi
\end{align}
the first gap equation yields ${\Delta_\uparrow=\Delta -U}$, and the order parameters are given by
\begin{align}
\begin{split}
\delta\rho_c=\frac{1}{2}+\Delta_\uparrow\kappa_\uparrow K(\kappa_\uparrow), \\
\delta\rho_s=\frac{1}{2}-\Delta_\uparrow\kappa_\uparrow K(\kappa_\uparrow).
\end{split}
\end{align}
Therefore ${\delta\rho_c+\delta\rho_s=1}$, which implies ${n_{i\downarrow}=1}$ on odd sites and 0 on even sites. We have verified, using both analytical and numerical tools, that this mean-field solution reproduces the exact ground state of this ionic Falicov-Kimball model, namely the immobile particles occupy the potential minima and the mobile particles experience an effective ionic potential of strength ${\Delta-U}$.
Both for ${U<\Delta}$, where ${\delta\rho_c>\delta\rho_s}$, and for ${U>\Delta}$, where ${\delta\rho_s>\delta\rho_c}$, the gap parameter $\Delta_\uparrow$ is finite, but for ${U=\Delta}$ it vanishes, and therefore the system is metallic just at the critical point ${U_c=\Delta}$. We can also easily show that the slopes of the order parameters diverge logarithmically by approaching the critical point. If $U$ tends to $U_c$ the parameter $\kappa_\uparrow$ tends to 1. Using the asymptotic behavior of the elliptic integrals together with Eq.~\eqref{eq:elliptic} we find
\begin{align}
\frac{d\delta\rho_s}{dU}=-\frac{d\delta\rho_c}{dU}\sim\log\frac{1}{1-\kappa_\uparrow}\sim\log\left|\frac{4}{U-U_c}\right|.
\end{align}

\section{Summary and outlook}\label{sec:sum}
In our study of the one-dimensional mass-imbalanced ionic Hubbard model we have found a subtle interplay of explicit and spontaneous symmetry breakings. We have explored in detail the competition between two types of order, namely a modulation of the particle density induced by the breaking of translational symmetry and alternating magnetic order originating from the broken SU(2) symmetry. We have limited ourselves to the ground state at half filling (one particle per site), using mostly mean-field theory. For given parameters ${\Delta>0}$ (ionic potential strength) and 
${t_\downarrow/t_\uparrow\neq 1}$ (hopping imbalance) we found a weakly first order transition as a function of the on-site interaction $U$ from a dominant density modulation to prevailing spin order. We have interpreted this transition in terms of an avoided criticality.
Tiny steps appear in the two order 
parameters at a point $U_c$. The transition is very clearly seen
as an infinitely sharp peak in the fidelity susceptibility.

Due to the broken symmetries, spin-dependent band gaps $\Delta_\uparrow$ and $\Delta_\downarrow$ appear which behave very differently as functions of $U$. At $U_c$ the gap for the light particles becomes very small, especially for large ionic potential strength 
$\Delta$, while the gap for the heavy particles remains large. This opens a very unusual avenue for spin-selected transport.

For symmetric hopping the existence of an intermediate phase is well established on the basis of advanced analytical and numerical methods.
We have verified that such an intermediate phase also appears in mean-field theory (for large values of $U$),
with a second-order transition at $U_{c1}$ and a (weakly) first-order transition at $U_{c2}$.
However, for an arbitrarily small mass imbalance the second-order transition is replaced by a crossover and the intermediate phase disappears.

Mean-field theory is quite generally expected to be a bad approximation for one-dimensional systems, but in the present case fluctuations are strongly suppressed due to pre-existing gaps in the charge and spin excitation spectra and we expect our phase diagram to be essentially correct. Our mean-field result for the ground-state energy agrees to leading order in $U$ with perturbation theory, and it also reproduces the exact ground state in the Falicov-Kimball limit (${t_\downarrow\rightarrow 0}$). 

The mean-field parameters $\Delta_\uparrow$ and $\Delta_\downarrow$, which determine the order parameters and also the gaps in the excitation spectrum behave very differently as $U_c$ is approached. The gap parameter for more mobile particles, $\Delta_\uparrow$, becomes very small and jumps to a negative value at $U_c$, while $\Delta_\downarrow$ remains large. Interestingly, the larger the strength of the ionic potential $\Delta$, the smaller the minimum gap reached at $U_c$. This value also strongly decreases with increasing mass imbalance and tends to zero in the Falicov-Kimball limit.  

We have also discussed several limiting cases of the model, some of which have been thoroughly investigated in the recent past. In the limit of spin-independent hopping our mean-field results differ markedly from results obtained previously by large-scale numerical simulations. We attribute this discrepancy to the fact that the single-particle term of the Hamiltonian has ungapped spin excitations. In this case simple perturbation theory is not applicable and mean-field theory cannot be trusted.

For spin-dependent hopping, fluctuation effects are weak in most parts of the phase diagram. However, they may still be relevant close to the transition where one of the gap parameters is very small. Therefore it would be interesting to study the model in the region close to $U_c$ using more sophisticated methods, such as bosonization for small $U$ and the mapping onto a spin-1 model for large $U$ (as used in Ref. [\onlinecite{Tincani_09}] for the ionic Hubbard model). This would shed more light on the nature of this peculiar order-order transition and, in particular, on the avoided criticality. We could also learn how the intermediate phase found in the ionic Hubbard model is affected by a small mass imbalance.

\begin{acknowledgments}

M.S. and G.I.J  acknowledge support from the Georgian National Science
Foundation through the Grant FR/265/6-100/14. D.B. and G.I.J. also
acknowledge support from the Swiss National Science Foundation
through the Project IZK0Z2-160773.  L.J. acknowledges support
from the World Federation of Scientist through the National Stipend
Program.

\end{acknowledgments}

\appendix
\section{Critical behavior for the ionic Hubbard model}
Here we present the solution of the gap equations for the ionic Hubbard model for $U>U_{c1}$. We use the notation $\bar{\Delta}$ for the gap parameter at $U_{c1}$.
Expanding
$\Delta_\sigma\kappa_\sigma K(\kappa_\sigma)$ in powers of
$\eta_\sigma=(\Delta_\sigma-\bar{\Delta})/\bar{\Delta}$ we find
\begin{align}
\Delta_\sigma\kappa_\sigma
K(\kappa_\sigma)=\bar{\Delta}\left(\bar{\kappa}K(\bar{\kappa})+\sum_{n=1}^\infty
c_n\eta_\sigma^n\right),
\end{align}
where the first three coefficients are given by
\begin{align}
c_1&=\bar{\kappa}[K(\bar{\kappa})-E(\bar{\kappa})],\nonumber\\
\label{eq:expansion} c_2&=-\frac{\bar{\kappa}}{2}
[(1-\bar{\kappa}^2)K(\bar{\kappa})-(1-2\bar{\kappa}^2)E(\bar{\kappa})], \\
c_3&=\frac{\bar{\kappa}}{6}
[2(1-3\bar{\kappa}^2+2\bar{\kappa}^4)K(\bar{\kappa})-(2-11\bar{\kappa}^2+8\bar{\kappa}^4)E(\bar{\kappa})]
.\nonumber
\end{align}
It is convenient to introduce the dimensionless coupling constant
$g=U_{c1}/(2\pi t)$. The soft-mode condition (\ref{eq:sm}) then
implies $c_1=1/g$. Using Eq. (\ref{eq:gap_av}) we can rewrite the
gap equations (\ref{eq:gap}) for $U=U_{c1}(1+\varepsilon)$ as
\begin{align}
\begin{split}
\eta_\uparrow&=-(1+\varepsilon)\eta_\downarrow-g(1+\varepsilon)\sum_{n=2}^\infty
c_n\eta_\downarrow^n
-g\varepsilon\bar{\kappa} K(\bar{\kappa}), \\
\eta_\downarrow&=-(1+\varepsilon)\eta_\uparrow-g(1+\varepsilon)\sum_{n=2}^\infty
c_n\eta_\uparrow^n -g\varepsilon\bar{\kappa} K(\bar{\kappa}).
\end{split}
\label{eq:gap_diff}
\end{align}
Subtracting the two equations we get
\begin{align}
(\eta_\uparrow-\eta_\downarrow)\varepsilon+g(1+\varepsilon)
\sum_{n=1}^\infty
c_{n+1}\left(\eta_\uparrow^{n+1}-\eta_\downarrow^{n+1}\right)=0\,.
\label{eq:gap_dev}
\end{align}
We now use the factorization
\begin{align}
\eta_\uparrow^{n+1}-\eta_\downarrow^{n+1}=(\eta_\uparrow-\eta_\downarrow)\,
p_n(\eta_\uparrow,\eta_\downarrow),
\end{align}
where $p_n(x,y)$ are polynomials in $x$ and $y$, in particular
\begin{align}
\begin{split}
p_1(x,y)&=x+y, \\
p_2(x,y)&=\frac{1}{4}(x-y)^2+\frac{3}{4}(x+y)^2.
\end{split}
\end{align}
For $\eta_\uparrow\neq\eta_\downarrow$ Eq. (\ref{eq:gap_dev})
becomes
\begin{align}
0=\varepsilon+g\, (1+\varepsilon)\sum_{n=1}^\infty c_{n+1}\,
p_n(\eta_\uparrow,\eta_\downarrow).
\end{align}
The leading contributions in powers of
$\eta_\uparrow+\eta_\downarrow$ and $\eta_\uparrow-\eta_\downarrow$
are
\begin{align}
0=\varepsilon+g\,
\left[c_2(\eta_\uparrow+\eta_\downarrow)+\frac{c_3}{4}(\eta_\uparrow-\eta_\downarrow)^2\right]\,
.
\end{align}
A second relation is obtained by adding the two lines of Eq.
(\ref{eq:gap_diff}). The leading terms are
\begin{align}
\eta_\uparrow+\eta_\downarrow=-g\left[\varepsilon\bar{\kappa}
K(\bar{\kappa})
+\frac{c_2}{4}(\eta_\uparrow-\eta_\downarrow)^2\right].
\end{align}
The last two equations yield the desired relations linking the gap
parameters and $\varepsilon=(U-U_{c1})/U_{c1}$ slightly above the
critical point $U_{c1}$,
\begin{align}
\begin{split}
\eta_\uparrow+\eta_\downarrow&=\frac{c_3\, \bar{\kappa}K(\bar{\kappa})\, g-c_2}{c_2^2\, g-c_3}\, \varepsilon, \\
(\eta_\uparrow-\eta_\downarrow)^2&=\frac{4[1-c_2\, g^2\,
\bar{\kappa}K(\bar{\kappa})]}{g\, (c_2^2\, g-c_3)}\, \varepsilon.
\end{split}
\end{align}
The order parameters are closely linked to the gap parameters. To
leading order in $(U-U_{c1})$ we find
\begin{align}
\delta\rho_s&\approx\frac{\bar{\Delta}}{4\pi t}c_1(\eta_\downarrow-\eta_\uparrow) \nonumber \\
&\approx \frac{\bar{\Delta}}{U_{c1}}\left[\frac{1-c_2\,
g^2\bar{\kappa}K(\bar{\kappa})}{g\, (c_2^2\,
g-c_3)}\right]^\frac{1}{2}
\left(\frac{U-U_{c1}}{U_{c1}}\right)^\frac{1}{2} 
\label{eq:spin}
\end{align}
for the magnetic order and
\begin{align}
&\delta\rho_c(U)-\delta\rho_c(U_{c1}) \nonumber \\
&\ \ \approx\frac{\bar{\Delta}}{4\pi
t}
\left[c_1(\eta_\uparrow+\eta_\downarrow)+\frac{c_2}{2}(\eta_\uparrow-\eta_\downarrow)^2\right]\nonumber\\
&\ \ \approx \frac{\bar{\Delta}}{2U_{c1}}\, \frac{c_2+
g\bar{\kappa}K(\bar{\kappa})(c_3-2c_2^2\, g)}{c_2^2\, g-c_3}\,
\left(\frac{U-U_{c1}}{U_{c1}}\right) 
\label{eq:charge}
\end{align}
for the charge modulation.

\end{document}